\documentclass{article}
\usepackage{amsmath}
\usepackage{graphicx} 
\usepackage[margin=3cm]{geometry}
\usepackage{hyperref}

\title{Learning temporal structure of the input with a network of integrate-and-fire neurons}

\author{Lyudmila Kushnir, Sophie Den{\`e}ve\\ \small
GNT - LNC, Departement d'etudes cognitives, Ecole normale superieure, 
\\ \small INSERM, PSL Research University, 75005 Paris, France
\\ \small}

\begin{document}

\maketitle

\section*{Abstract}

The task of the brain is to look for structure in the external input. We study a
network of integrate-and-fire neurons with several types of recurrent connections that learns the
structure of its time-varying feedforward input by attempting to efficiently represent this input with
spikes. The efficiency of the representation arises from incorporating the structure of the input into
the decoder, which is implicit in the learned synaptic connectivity of the network. While in the
original work of [Boerlin, Machens, Den{\`e}ve 2013] and [Brendel et al., 2017] the structure learned
by the network to make the representation efficient was the low-dimensionality of the feedforward
input, in the present work it is its temporal dynamics. The network achieves the efficiency by
adjusting its synaptic weights in such a way, that for any neuron in the network, the recurrent
input cancels the feedforward for most of the time. We show that if the only temporal structure that the 
input possesses is that it changes slowly on the time scale of neuronal integration, the dimensionality of the network 
dynamics is equal to the dimensionality of the input. However, if the input follows
a linear differential equation of the first order, the efficiency of the representation can be increased by increasing 
the dimensionality of the network dynamics in comparison to the dimensionality of the input. 
If there is only one type of slow synaptic current in the network, the increase is two-fold, while if there are 
two types of slow synaptic currents that decay with different rates and whose amplitudes can be adjusted separately, it
is advantageous to make the increase three-fold. We numerically simulate the network with synaptic weights that imply the most efficient input representation in the above cases. We also propose a learning rule by means of which the corresponding synaptic weights can be learned. 

\section*{Introduction}

In this work we generalize and give a new interpretation of the neural encoding scheme suggested in \cite{Boerlin2013} and further developed in \cite{LongSI},\cite{Bourdoukan2015},\cite{Alemi2017}, which considers neural networks of integrate-and-fire neurons that perform one of the two tasks. The first task is encoding the feedforward input by the time sequence of the spikes of the network and the second one is encoding the solution of a particular differential equation with the source term given by the feedforward input. In the present work we focus on the first task, which we refer to as autoencoding. The framework in question provides an account for the trial-to-trial variability ubiquitously observed in the neural systems, which is an alternative to modeling the neurons as Poisson spike generators. This framework also assumes an advantageous scaling of the accuracy of the representation with the number of neurons, in comparison to Poisson units. 

Even though these are among the strongest points of the discussed framework,  we don't consider them to be the only ones. 
We propose that the model suggests a more general perspective on the operation of the neural system that spans different levels of organization. To find such perspective means to understand in a unified way not only how the nervous system solves particular computational tasks, but also how these tasks arise, how it "knows" that these computations should be performed. 

We suggest that studying an efficient way to encode a structured input in a neural network with the time sequence of the spikes of its neurons is a good place to start in searching for this unifying perspective. The reason is that in order to be efficient, the network should be able to extract the existing structure from the input and represent only the information that can not be predicted based on knowing this structure. To clarify, we use the term "structure" in the most general sense, to mean any form of predictability in the input. 

In the current work we will demonstrate, based on \cite{Boerlin2013} and \cite{LongSI}, that extracting the existing structure of the input is equivalent to
adjusting the synaptic efficacies in the way as to keep membrane potentials of all the neurons close to the rest value whenever possible and adjusting either neuronal gains or firing thresholds to keep the same the average activity of all the neurons. This relates an objective of information encoding, which is a task assigned to the network by a researcher, to something that is formulated in purely network terms without appealing to any externally imposed task. 

This observation is in line with seeing the goal to maintain the homeostasis as the main principle of functioning of the organism at various levels of organization. The attractiveness of this hypothesis is that it seems to be plausible in the context of evolution, offering a scenario of gradual complexification of life by adding new links connecting different components, such as different parts of the organisms body, different brain areas or connecting the organism to different aspects of the environment by means of new sensory organs or modification of the old ones. With more links added, the task of maintaining homeostasis whenever possible requires discovering more and more complex structures in the environment, so that the corresponding homeostatic changes in the organism can be predicted and compensated for to keep the homeostatic state. Very abstractly put, in this framework the organism is trying to autoencode its own response to complex environment, and all other computational tasks can be seen as arising as "obstacles" on the way. 

The model presented in this paper is formulated on the network level assuming the simplest dynamics for an individual unit: integrate-and- fire neuron (IF neuron). We point out, however, that this dynamics can itself be viewed as realization of the same principle that we propose for the operation of the network (see section \ref{sec:Balance}). The network is receiving a time-varying feed-forward input and its task is to represent this input in real time with as few spikes as possible, so that it can be easily decoded from the network activity. What makes the representation efficient, in the sense of using less spikes, is incorporating the existing structure of the input, such as low dimensionality \cite{Boerlin2013} or temporal  predictability (sections \ref{sec:SB}, \ref{sec:TS}) into the network. 

The central principle that allows to build a network corresponding to a particular input structure is the principle of feedforward-recurrent balance (section \ref{sec:Balance}). In a nutshell, it can be formulated as adjusting the recurrent connections in the network in such a way as to make the recurrent contribution to the neuronal voltages cancel the feedforward contribution, so that the voltages are maintained close to the rest values whenever possible. As the recurrent contribution can be unambiguously recovered from the network activity by an external decoder, the feedforward contribution can also be decoded. This observation implies that maintaining the balance of the voltage is equivalent to faithfully encoding the feedforward input. The efficiency of the representation is improved when the structure of the feedforward input can be reproduced by the recurrent, as this improves the feedforward-recurrent balance and delays the next population spike. 

The flexibility in the temporal profile of the synaptic currents determines what kind of structure present in the feedforward input can be incorporated into the network. The network with only fast connections, described in \cite{Boerlin2013} and in the section \ref{sec:fastOnly}, can incorporate low-dimensional but not temporal structure of the input. A network with a slow synaptic current whose temporal profile is fixed (see section \ref{sec:SB}) is able to match the feedforward current at the time of the spike, if in addition to being low dimensional the feedforward input changes slowly in time. This improves the balance and makes the representation more efficient. The network with variable temporal profile of the synaptic current, which we model by introducing two types of slow currents that decay with different time constants and whose relative strengths can be adjusted locally (see section \ref{sec:DE2}), can match not only the input, but also its first time derivative at the time of the spike, which further increases the efficiency of the representation. This is only possible, however, if the input possesses the temporal structure of following a homogeneous first order differential equation, which implies that its evolution can be predicted from the current value. Note, that when the feedforward input stops obeying the differential equation, it is still represented faithfully by the network's spikes, only this representation is less efficient, which is immediately reflected in the increase of the firing rate of the network. 

Our model can also be seen as a network realization of the predictive coding framework \cite{Spratling2017review}, where the expectation about the feedforward input is implemented in the value or temporal evolution of the recurrent input and is updated at every spike according to the structure of the network connectivity.  

Although we do not investigate in detail in this work how this connectivity structure can be learned, we do propose in the Discussion (section \ref{sec:Dis}) an unsupervised learning rule, inspired by \cite{LongSI}. This learning rule is local and follows from the aim of keeping the postsynaptic voltage at zero after a presynaptic spike. We hypothesize that this learning rule will lead to the network incorporating the structure present in the input most of the time. We also discuss briefly a possible generalization of the presented framework, in which the recurrent input that implements the prediction is generated not in the synapses but in a separate network, that we call a simulator network. The advantage of the two-subnetwork system is that the possible predictions that can be realized are not constrained by the synaptic dynamics and can, for example, be on a much longer time scale. Learning in such a system can happen at two levels. At the first level, the connections between the two subnetworks are tuned in such a way as to initialize the fixed dynamics of the simulator to best match the prediction about the feedforward input. This is conceptually similar to learning of the synaptic connections with the aim of maintaining the feedforward-recurrent balance in a single network. When a particular temporal pattern of the feedforward input persists, or is considered important for other reasons, the connections within the simulator network can be changed, so that the dynamics of the simulator matches the input dynamics \cite{LearningSlowConnections}. This potentially can be a model of learning transfer between different brain areas.  

In the light of spike-coding versus rate-coding debate, our framework takes an intermediate stand. From purely decoding point of view, it can be interpreted as the rate code, since the estimate of the encoded variable is a convolution of the spike train with a smooth kernel (or a sum of such convolutions). Changing a timing of a particular spike gradually degrades the decoding performance. From the point of view of network dynamics, however, the model is not a rate model: it is deterministic and the time of every spike can be predicted exactly, given the feedforward input into the network. Moreover, every spike is interpretable, but instead of neurons being selective to a particular feature of the input, in our framework the identity of the neuron that has just spiked determines unambiguously the current representation error. In the case of the network that has learned a temporal structure of the input, neurons are selective to a combination of the current representation error and the expected future error, namely the difference between the future evolution of the recurrent input and the prediction about the evolution of the feedforward one. The trial-to-trial variability in our model arises from a the network being extremely sensitive to the smallest variations in the input. 

We also point out that the function of the fast connections in the model network presented here (section \ref{sec:fastOnly}) is the same as in the original work \cite{Boerlin2013}, while the slow connections described in the sections \ref{sec:SB} and \ref{sec:TS} are conceptually different from the slow connections of \cite{Boerlin2013}. In the model of \cite{Boerlin2013}, the slow connections are introduced in order to make the network produce the output which is a solution of a differential equation with the source term given by the feedforward input. In the present work, on the other hand, the feedforward input itself is following a differential equation, and the job of the slow connections is to incorporate this temporal structure into the network to make it more efficient in encoding the input. 

\section{Network setup}\label{sec:1}
Our network consists of $N$ leaky integrate-and-fire neurons, that are receiving a time-varying feedforward input $I^{ext}$ and are coupled by means of several types of synaptic currents $h^{(a)}$ with corresponding connectivity matrices $\Omega^{(a)}$. By discriminating between different types of synaptic currents, we imply different 
dynamics of these currents after a presynaptic spike. A larger variety of synaptic currents makes the network more flexible and better in discovering a temporal structure of its feedforward input, as will become clear later.

The membrane voltages $V_i$ of the network are governed by the following dynamics:
\begin{align}
&\dot V_i(t) = -\lambda V_i(t) + I^{ext}_i(t) + \sum_{a}\sum_{j = 1}^N\Omega^{(a)}_{ij}h^{(a)}_j(t) \hspace{1cm} \text{for  } i = 1\dots N\label{eq:IF} \\
&V_i(t) = T_i\hspace{1cm} \text{neuron  }i \text{  spikes and    }  V_i\rightarrow 0  \notag
\end{align} 

The reset mechanism of the integrate-and-fire neuron can be incorporated into equation (\ref{eq:IF}) by adding a delta-function contribution to the input current of a neuron at the moment of its spike:
\begin{align}
&\dot V_i(t) = -\lambda V_i(t) + I^{ext}_i(t) + \sum_{a}\sum_{j = 1}^N\Omega^{(a)}_{ij}h^{(a)}_j(t) - T_i \sum_{s = 1}^{N_{i}}\delta(t-t_s^i)\hspace{1cm}\notag \\
&\text{where } t_s^i \text{ are such that}\\
&V_i(t_s^i) = T_i\hspace{1cm} 
\label{eq:IF1}
\end{align}
Here $N_i$ is the total number of spikes of the neuron $i$, $t_s^i$ is the time of the spike $s$ of neuron $i$, and $\delta(t-t_s^i)$ is the Dirac delta-function.
 
Assuming the spike times are known, this equation can be solved, leading to:
\begin{align}
&V_i(t)  = \hat I_i^{ext}(t) + \sum_a\sum_{j = 1}^N\Omega^{(a)}_{ij}\hat h^{(a)}_j(t) - T_i r_i(t)\label{eq:IFI}\\
&V_i(t_s^i) = T_i\hspace{1cm} \text{:  spike of neuron  }i  \notag
\end{align}
where 
\begin{align} 
&\hat I_i^{ext}(t) = \int_{-\infty}^t I_i^{ext}(t')\text{e}^{-\lambda(t-t')}dt'\label{eq:hatI}\\
&\hat h^{(a)}_i(t) = \int_{-\infty}^t h^{(a)}_i(t')\text{e}^{-\lambda(t-t')}dt'\label{eq:hath}\\
&r_i(t) = \int_{-\infty}^t \sum_{s = 1}^{N_i}\delta(t'-t_s^i)\text{e}^{-\lambda(t-t')}dt'= \sum_{s = 1}^{N_i}\text{e}^{-\lambda(t-t_s^i)}\label{eq:hatr}
\end{align}
are corresponding inputs filtered with the exponential decay kernel $\text{e}^{-\lambda t}$.

We will always assume that the typical time scale of the evolutions of the external input $|I_i^{ext}/\dot{I}_i^{ext}|$ is slow in comparison to the membrane time scale $\frac{1}{\lambda}$. In this case,  $\hat I^{ext}_i(t)$ is approximately proportional to $I^{ext}_i(t)$ with the rescaling factor $\frac{1}{\lambda}$:
\begin{equation}
\left|\frac{\dot{I}_i^{ext}}{I_i^{ext}}\right|\ll \lambda \hspace{0.5cm}\implies\hspace{0.5cm} \hat I_i^{ext} \approx \frac{1}{\lambda}I_i^{ext}
\label{eq:Iscale}.
\end{equation}

We also assume that the feedforward input is large $\hat I^{ext}_i\gg T_i$, so that the neurons are in the input-driven regime and in the absence of recurrent contribution would have spiked regularly with a high rate. 

\section{Feedforward-recurrent balance and efficiency} \label{sec:Balance}
In this work we assume that the task of the network is to represent its time-varying feedforward input with its spikes. 
Following \cite{Boerlin2013} we define representation efficiency as the number of spikes that the network spends to represent its input given an upper bound on the representation error. The central idea of the present framework, introduced in \cite{Boerlin2013} and further clarified in \cite{LongSI}, is that balancing the feedforward input by the recurrent contribution, which keeps the neuronal voltages close to zero 
\begin{equation}
V_i(t)\approx 0 \hspace{1cm} \text{for  } i = 1\dots N,
\end{equation}   
is equivalent to the efficiency of input representation. 

Indeed, if the time-varying feedforward input is approximately balanced by the recurrent, the recurrent input reconstructed from the network's spikes and multiplied by $-1$ gives an online estimate of the feedforward input. To guarantee an upper bound on the representation error, we need to guarantee the precision of the balance. In other words, as long as the balance is violated beyond the tolerance margin, the network should spike to correct the recurrent input, restoring the balance and updating the estimate. 

As integrate-and-fire neurons, considered here, spike based on their voltages, not their inputs, the upper bound on the error can only be set for representing the filtered version of the input $\hat I_i^{ext}(t)$, not the input itself (see (\ref{eq:IFI})). However, if the input is assumed to be slow, $\hat I_i^{ext}(t)$ and $I_i^{ext}(t)$ differ only by a scale (see (\ref{eq:hatI}) and (\ref{eq:Iscale})). We will loosely refer to the representation of the leaky integral of the input $\hat I_i^{ext}(t)$ by the network as input representation, keeping in mind that input fluctuations on the time scale smaller than $\frac{1}{\lambda}$ can not be represented by the suggested scheme. 

The reset mechanism of an isolated integrate-and-fire neuron by itself can be seen as enforcing the feedforward-recurrent balance if the self-inhibition after a spike is considered as a recurrent delta-function input (see \cite{LongSI}). When the voltage of the neuron reaches the threshold, a spike is fired conveying this information to the decoder. For a single neuron, the analogue of the equation (\ref{eq:IFI}) is
$$
V(t) = \hat I^\text{ext}(t) - Tr(t)
$$
where $r(t)$ is the spike train of the neuron filtered with the exponential kernel, as given in (\ref{eq:hatr}). We interpret $Tr(t)$ as the estimate of $\hat I^\text{ext}(t)$, which makes the voltage $V(t)$ play the role of the representational error. As $V(t)$ is bounded by the firing threshold from above, but not from below, the representational error can be controlled only if $I^\text{ext}(t)$ is constraint to be positive ($\hat I^\text{ext}(t) - Tr(t)$ can not become negative), in which case the error can not exceed the threshold $T$. What follows can be seen as an extension of this encoding scheme to the case of multi-dimensional input with different neurons representing different directions in the space of inputs. 

Coming back to the network of $N$ neurons (\ref{eq:IFI}) and assuming no recurrent connections ($\Omega_{ij} = 0$), we see that the filtered input into the neuron $i$, $\hat I^\text{ext}_i(t)$, is represented by the spikes of this neuron as $\hat I_i(t)\approx T_ir_i(t)$ if $I^\text{ext}_i(t)$ is always positive. However, when the input has a low-dimensional structure, namely the vector $\boldsymbol{I^\text{ext}}(t)$ changes in a $J$-dimensional subspace of the $N$-dimensional space and $J\ll N$, introducing connections between neurons ($\Omega_{ij}\neq 0$) can make the repesentation much more efficient. Namely, the total number of population spikes required for the same upper bound on the representational error can be decreased by a factor of $\frac{\sqrt{2}}{N}$ in the limit $N\rightarrow\infty$, $J = \text{const}$. Also, the constraint on $I_i^\text{ext}(t)$ being positive can be lifted. 

In the next section we will show that for strictly low-dimensional input, a spike of one neuron in the network can bring not only its own voltage, but also the voltages of all other neurons to zero, thus delaying the next population spike. In order to achieve this, the neurons should interact by means of synaptic currents that are large, but short-lasting on the time scale of neuronal integration. We will describe this mechanism in detail and derive the required connectivity matrix, following \cite{Boerlin2013}. If the input is approximately low-dimensional, the proposed connectivity structure will still reduce the number of spikes to the extent dependent on the ratio of the low-dimensional and the full-dimensional parts of the input. 

If, in addition to low-dimensionality of the input, the future input can be predicted based on the current state of the network, the voltages of all the neurons can be kept close to zero for some period after a population spike. In this case, the next spike will be delayed even further. We will discuss this scenario in the following sections.  

In summary, the spatial (low dimensionality) or temporal structure of the feedforward input can be incorporated into the network connectivity to 
enhance the balance, which decreases the number of spikes fired by the network, without sacrificing the accuracy with which the feedforward input can be reconstructed from the network's spikes. 

\section{Neuron-to-neuron predictability, fast connections}\label{sec:fastOnly}
In this section we describe how a network of integrate-and-fire neurons can exploit the low-dimensional structure of its input to represent it more efficiently, compared to population of non-connected neurons. This network is a slight modification of the autoencoder of \cite{Boerlin2013} and \cite{LongSI}, looked at from a somewhat different perspective. 

We assume that the feedforward input into the network $\boldsymbol{I^{ext}}(t)$ varies in a subspace spanned by the columns of an non-degenerate $N\times J$ - dimensional matrix $\boldsymbol{F}$. This is equivalent to saying, that there are $J<N$ time-varying signals $c_\alpha(t),\hspace{0.1cm}\alpha = 1\dots J$ 
and 
\begin{equation}
I_i^{ext}(t) = \sum_{\alpha=1}^JF_{i\alpha}c_\alpha(t)\hspace{1cm} i = 1\dots N
\label{eq:Ilow}
\end{equation}
 
We refer to the $i$-th row of the matrix $\boldsymbol{F}$ as the vector of feedforward weights of the neuron $i$. The feedforward input into a neuron $i$ is than given by an inner product of two $J$-dimensional vectors: $\boldsymbol{c}(t)$ and $\mathbf{F_i}$. 

We scale the feedforward input in such a way that the Euclidean norm of $\boldsymbol{c}(t)$ is of the order of neuronal integration constant $\lambda$, 
$$
|\boldsymbol{c}(t)|=O(\lambda).
$$ 
This implies that the leaky integral $\boldsymbol{\hat c}(t)$ of the input, which is given by
\begin{equation}
\boldsymbol{\hat c}(t)= \int_{-\infty}^t \boldsymbol{c}(t')\text{e}^{-\lambda(t-t')}dt'
\label{eq:chat}
\end{equation}
and is the variable represented by the network, is of order one:
$$
|\boldsymbol{\hat c}(t)| = O(1)
$$

At this point we don't impose any additional assumptions on the input $\boldsymbol{c}(t)$.
 
In this section we consider the network with only one type of synaptic current, namely \emph{fast current} $\boldsymbol{h^\text{f}}$ which we assume to be short-lasting on the scale of neuronal time constant $\frac{1}{\lambda}$.
 
\begin{align}
&\dot V_i(t) = -\lambda V_i(t) + \sum_{\alpha=1}^JF_{i\alpha}c_\alpha(t) + \sum_{j = 1}^N\Omega^{\text{f}}_{ij}h^{\text{f}}_j(t) - T_i \sum_{s = 1}^{N_{i}}\delta(t-t_s^i)\hspace{1cm}\notag \\
&V_i(t_s^i) = T_i\hspace{1cm} \text{:  spike of neuron  }i 
\label{eq:IF30}
\end{align}

The postsynaptic current of neuron $i$, $h^\text{f}_i(t)$ is the sum of synaptic kernels $o(t-t_s^i)$ placed at the times of the neuron's spikes
\begin{equation}
h^\text{f}_i(t) =  \sum_{s = 1}^{N_{i}}o(t - t_s^i).
\label{eq:ho}
\end{equation}
We assume that the synaptic kernel $o(t-t_s^i)$ has support $\Delta t$ and is normalized as
\begin{equation}
\int_0^{\Delta t} o(t')\text{e}^{-\lambda(\Delta t - t')}dt' = 1
\end{equation}
 (the normalization factor can be absorbed in the connectivity matrix). 
 We also assume that $\Delta t$ is small when compared to the neuronal time constant $1/\lambda$
$$
\Delta t\ll\frac{1}{\lambda}
$$
which together with $|\boldsymbol{c}(t)|=O(\lambda)$ implies that the synaptic kernel $o(t-t_s)$ can be substituted by Dirac delta function: 
\begin{equation}
h_i^\text{f}(t) = \sum_{s = 1}^{N_{i}}\delta(t - t_s^i)
\label{eq:hf}
\end{equation}

Then, the equation (\ref{eq:IF30}) becomes
\begin{align}
&\dot V_i(t) = -\lambda V_i(t) + \sum_{\alpha = 1}^JF_{i\alpha}c_\alpha(t) + \sum_{j = 1}^N\Omega^\text{f}_{ij}\sum_{s = 1}^{N_j}\delta(t-t_s^j)
 \label{eq:IF3}\\
&V_i(t_s^i) = T_i\hspace{1cm} \text{:  spike of neuron  }i \notag
\end{align}
where we have absorbed the last term in (\ref{eq:IF30}) into the recurrent interaction by redefining the diagonal entries of $\Omega^\text{f}$.

We can now integrate (\ref{eq:IF3}) to get
\begin{align}
&V_i(t) = \sum_{\alpha = 1}^JF_{i\alpha}\hat c_\alpha(t) + \sum_{j = 1}^N\Omega^\text{f}_{ij}r_j(t)\label{eq:IF2}\\
&V_i = T_i\hspace{1cm} r_i\rightarrow r_i +1\notag
\end{align}
where 
\begin{equation}
\hat c_\alpha(t) = \int_{-\infty}^tc_\alpha(t')\text{e}^{-\lambda(t-t)}dt',
\end{equation}
and $r_i$ is given by (\ref{eq:hatr}). 

Following the idea described in the previous section, we impose the feedforward-recurrent balance by choosing the connectivity matrix $\Omega^\text{f}$ in such a way that all the voltages are close to zero for as much time as possible. Since the synaptic currents are short-lasting, the best that the network can do is to bring all the voltages to zero after every population spike and let them evolve following on the feedforward input between the spikes. We note that the proposed scheme only  implements the greedy minimization of $\boldsymbol{V}^2(t)$ and is not an optimal solution to minimizing $\int\boldsymbol{V}^2(t)dt$. 

The network's ability to bring the voltages of all the neurons to zero after a spike of a particular neuron is surprising when the dimensionality $J$ of the input is greater than one. Indeed, the fact that one of the neurons has reached its threshold should give the full information about the voltages of all other neurons at this time. In other words, the voltage of a neuron $j$ should be the same every time a given neuron $i$ spikes. We will now demonstrate, that this is indeed the case under certain conditions.  

We assume that the fast connectivity matrix $\Omega^\text{f}$ is of the form 
\begin{equation}
\Omega^\text{f}_{ij} =  - \sum_{\alpha = 1}^JF_{i\alpha}D^\text{f}_{j \alpha}\label{eq:flr}
\end{equation}
Namely, it has a low rank structure with the left singular vectors spanning the same subspace as the columns of the feedforward matrix $\boldsymbol{F}$ ($\text{Im}\boldsymbol{\Omega^\text{f}} = \text{Im}\boldsymbol{F}$). We call this subspace \emph{$F$-space}. 

In this case, the vector of voltages of all the neurons is given by (see (\ref{eq:IF2}))
\begin{equation}
V_i(t) = \sum_{\alpha = 1}^JF_{i\alpha}\hat d_\alpha(t)
\label{eq:Vd}
\end{equation}
where we have introduced a $J$- dimensional vector 
\begin{equation}
\hat d_\alpha(t) = \hat c_\alpha(t) - \sum_{j = 1}^N D^\text{f}_{j \alpha}r_j(t)
\label{eq:dhat}
\end{equation}
We call $\boldsymbol{\hat d}(t)$ \emph{state vector}, as it determines the internal state of the network unambiguously by (\ref{eq:Vd}), and the $J$-dimensional space of all possible values of $\boldsymbol{\hat d}$ - \emph{state space} of the network. 

The time evolution of the state vector is given by
\begin{equation}
\dot{\hat d}_{\alpha}(t) = -\lambda\hat d_\alpha(t) + c_\alpha(t) - \sum_{j = 1}^ND^\text{f}_{j \alpha}\sum_{s = 1}^{N_j}\delta(t - t_s^j)\label{eq:evdhat}
\end{equation}
which is just the equation of time evolution of the vector of voltages (\ref{eq:IF3}) multiplied by the pseudoinverse of the matrix $\boldsymbol{F}$.

\begin{figure}
\includegraphics[width = 13cm]{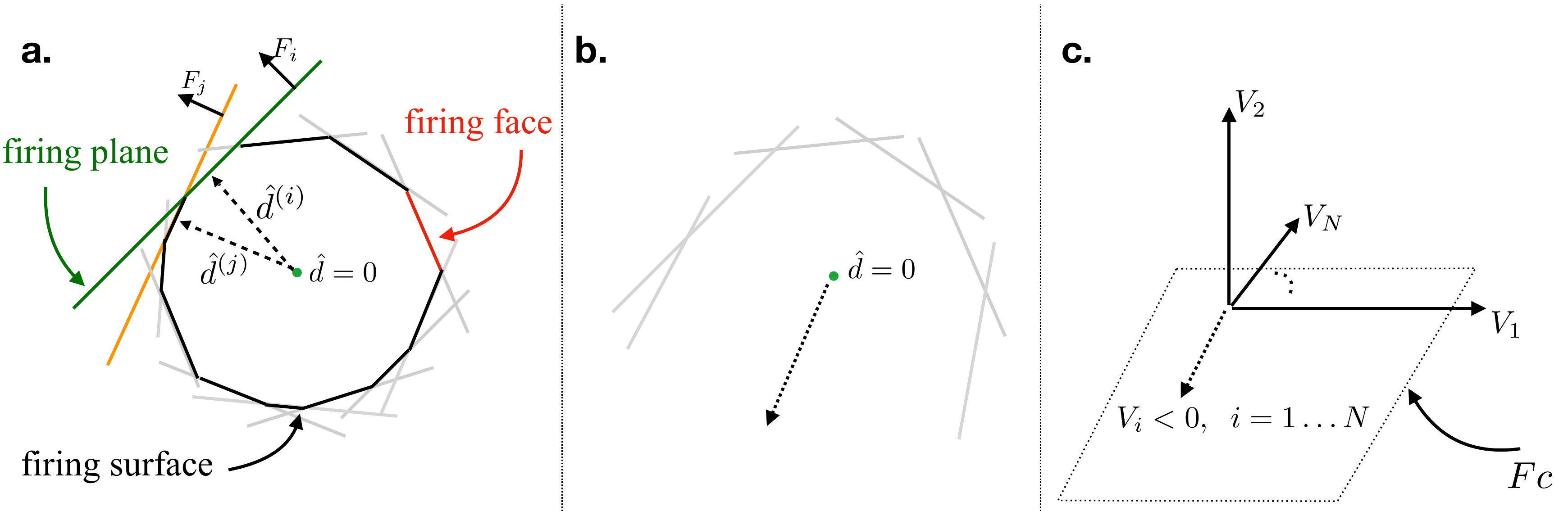}
\centering
\caption{\textbf{a.} Firing surface for a network of $N = 11$ neurons. \textbf{b.} Open box \textbf{c.} The case of the open box in neuronal space. }
\label{fig:box}
\end{figure}

A given neuron $i$ spikes when its voltage reaches the threshold $T_i$, 
\begin{equation}
V_i = \sum_{\alpha = 1}^JF_{i\alpha}\hat d_{\alpha} = T_i\label{eq:ispikes}
\end{equation}
This condition describes a $(J-1)$-dimensional hyperplane in the $J$-dimensional state space. We refer to this hyperplane as \emph{firing plane} of the neuron $i$. The orientation of the firing plane $i$ is determined by the vector $\boldsymbol{F_i}$ of feedforward weights of the neuron, while the Euclidean distance from the origin $\boldsymbol{\hat d} = \boldsymbol{0}$ to the firing plane is given by $\frac{T_i}{|\mathbf{F_i}|}$, with $|\mathbf{F_i}|$ representing the Euclidean norm of the vector $\mathbf{F_i}$ (see figure \ref{fig:box}a). We choose the firing thresholds $T_i$ in such a way, that this distance is the same for all neurons in the network and is equal to $\omega$ which, as we will see, determines the accuracy of the input representation:
\begin{equation}
T_i = \omega |\mathbf{F_i}|\hspace{1cm}\text{for } i = 1\dots N
\label{eq:sph}
\end{equation}
When the state vector $\mathbf{\hat d}(t)$ reaches one of the firing planes, the corresponding neuron $i$ fires a spike, increasing $r_i$ by 1 (see (\ref{eq:IF2})) and changing $\mathbf{\hat d}(t)$ by $-\mathbf{D^\text{f}_i}$, where $\mathbf{D^\text{f}_i}$ is the corresponding column of the matrix $\mathbf{D^\text{f}}$ (see (\ref{eq:dhat})). This change happens in a step-wise manner because we have approximated large and short-lasting synaptic currents by $\delta$-functions. 

We call the intersection of the half-spaces 
 $$
\boldsymbol{F_i}\boldsymbol{\hat d} \leq T_i\hspace{0.5cm} i = 1\dots N
$$
\emph{state domain} ${\cal{D}}$. This is the region of the state space where none of the neurons in the network fires. We refer to the boundary of this region ${\cal{D}}$ as \emph{firing surface}. 

The state domain can be either a convex polyhedron as on the figure \ref{fig:box}a, or have an open end as is in \ref{fig:box}b. The latter case is equivalent to the input subspace overlapping with the sector $V_i < 0,\quad i = 1\dots N$, so that for some directions of $\boldsymbol{c}$ all the neurons receive hyperpolarizing input and no neuron spikes even if the norm of $\boldsymbol{c}$ is large (figure \ref{fig:box}c).   

When the angles between any pair of neighboring faces of the firing surface are greater or equal to 90 degrees, which we will always assume, it is possible to choose the matrix $\boldsymbol{D^\text{f}}$ in such a way, that the state vector $\boldsymbol{\hat d}(t)$ never leaves the state domain. So, the dynamics of the network is described by the vector $\boldsymbol{\hat d}(t)$ evolving according to 
$\boldsymbol{\dot{\hat{d}}} = -\lambda\boldsymbol{\hat d}(t) + \boldsymbol c(t)$ inside the state domain and reflecting from the firing surface. 

The image of a vector bouncing within a polyhedron shape was introduced in \cite{NatureNeuroscience} for illustrative purposes and used in \cite{Nuno} to investigate the stability of the efficient balanced network. We point out that in these studies, the vector that is confined to stay within the polyhedron is the error of the representation of the feedforward input by the spikes of the network, while in the current work it is the state vector $\mathbf{\hat d}(t)$. The two are mathematically equivalent to each other, however the formulation of  \cite{NatureNeuroscience} and \cite{Nuno} implies representational framework, while the current derivation is purely dynamical.  

In general, spike of a particular neuron conveys the information only about the projection of the state vector $\mathbf{\hat d}(t)$ onto the vector of its input weights (\ref{eq:ispikes}). In other words, the spike means that the instantaneous value of $\mathbf{\hat d}(t)$ is somewhere on the firing plane of the spiking neuron,  which is not enough to specify the voltages of other neurons in the network. However, not any point on a given firing plane can be reached from the origin $\mathbf{\hat d} = \mathbf{0}$ without crossing a firing plane of another neuron in the network. We refer to the segment of a firing plane of neuron $i$ that can be reached from the origin as \emph{firing face} of neuron $i$ (see figure \ref{fig:box}). If the firing face is small, the neuron will spike only when the state vector $\mathbf{\hat d}(t)$ is close to a particular value, which implies one-to-one correspondence between the identity of the spiking neuron and the instantaneous value of the state vector.

The firing faces of all neurons will be small if the number of neurons $N$ is much larger than the dimensionality of the state space $J$, and the weight vectors $\boldsymbol F_i$ sweep all the directions approximately uniformly. In the neuronal space this condition corresponds to the fact that any direction of the input subspace ($\text{Im}\boldsymbol{F}$) has positive component along many neuronal axes, so that a small change in this direction will change the identity of the neuronal axis along which the component is largest. 

The spike of a neuron $i$ at time $t_s^i$ implies that the instantaneous value of the state vector $\boldsymbol{\hat d}(t_s^i)$ is somewhere on the firing face $i$, which can be written as 
\begin{equation}
\mathbf{\hat d}(t_i) = \mathbf{\hat d^{(i)}} + \mathbf{\Delta_s^i},
\end{equation}
where we have introduced the \emph{preferred vector} of the neuron $i$, $\mathbf{\hat d^{(i)}}$, which is the normal drawn from the origin onto the firing plane $i$ (see figure \ref{fig:box}):
\begin{equation}
\mathbf{\hat d^{(i)}} = \omega\frac{\mathbf{F_i}}{|\mathbf{F_i}|} =  \frac{T_i}{\mathbf{F_i}^T\mathbf{F_i}}\mathbf{F_i}
\label{eq:di}
\end{equation}
The preferred vector of the neuron is the vector of principal components of the collective voltage of the network $\boldsymbol{V}$ to which the neuron is selective. 

The difference $\boldsymbol{\Delta_s^i}$ between the value of the state vector at the moment of a spike and the preferred vector of the spiking neuron is bounded by the diameter of the firing face $i$. The information about the value of $\boldsymbol{\Delta_s^i}$ is not conveyed to the decoder. 

We introduce a bound $\Delta$ on the size of the firing faces of all the neurons in the network. To define $\Delta$ we first define the radius of a firing face $i$ as the distance from the base of the normal $\mathbf{\hat d^{(i)}}$ to the furthest point on the firing face. Then $\Delta$ is the maximum of these radii over all firing faces, so that 
\begin{equation}
|\mathbf{\Delta_s^i}|\leq \Delta \hspace{0.5cm}\forall i,s\label{eq:Delta}
\end{equation}
For fixed input dimensionality $J$, if the directions of the feedforward weights $\boldsymbol{F_i}$ are chosen to cover the sphere in the $J$-dimensional space uniformly, $\Delta \propto \frac{1}{N}$.

Since a spike of a neuron $i$ changes the state vector $\boldsymbol{\hat d}(t)$ by $-\mathbf{D^\text{f}_i}$,
in the infinitely large network, the state vector $\boldsymbol{\hat d}(t)$ can be brought to zero after every spike by defining the $i$-th column of the matrix $\boldsymbol{D^\text{f}}$ as
\begin{equation}
\mathbf{D^\text{f}_i} = \mathbf{\hat d^{(i)}} = \frac{T_i}{|\mathbf{F_i}|^2}\mathbf{F_i} = \omega\frac{\mathbf{F_i}}{|\mathbf{F_i}|}
\label{eq:fastD}
\end{equation}
This implies that the vector of neuronal voltages $\boldsymbol{V}(t)$ is also zero after every population spike. 
\begin{align}
\boldsymbol{\hat d}(t_s^{i+}) = 0\notag\\
\boldsymbol{V}(t_s^{i+}) = 0\notag
\end{align}

In the network with a finite number of neurons, and thus a finite $\Delta$, the state vector and the vector of neuronal voltages are only guaranteed to be close to zero after every spike, more precisely 
\begin{align}
|\boldsymbol{\hat d}(t_s^{i+})| \leq \Delta\notag\\
|\boldsymbol{V}(t_s^{i+})| \leq f\Delta\notag
\end{align}
where $f$ is the largest singular value of the matrix $\boldsymbol{F}$ (square root of the largest eigenvalue of $\boldsymbol{F}^T\boldsymbol{F}$). 

When the state domain is closed and the connectivity matrix is given by (\ref{eq:flr}) and (\ref{eq:fastD}), the dynamics of the network is equivalent to a point moving in the $J$-dimensional box of figure \ref{fig:box}a, which bounces back to zero (approximately, for a finite-size network) every time it hits a wall. 

As the state vector $\mathbf{\hat d}(t) = \boldsymbol{\hat c}(t) - \boldsymbol{Dr}(t)$ is bounded by this box (see (\ref{eq:dhat})), the leaky integral of the input $\mathbf{\hat c}(t)$ can be read out from the spikes of the network with a decoder $\mathbf{D}$ applied to the filtered spikes of the network $\mathbf{r}(t)$ 
\begin{equation}
\hat c(t)_\alpha \approx  \sum_{j = 1}^ND^\text{f}_{j \alpha}r_j(t)
\end{equation}

The error of this representation is bounded by the linear size of the state domain $\cal{D}$ (the furtherst point in $\cal{D}$ from the origin). In the limit of infinite network,
\begin{equation}
|\mathbf{\hat c}(t) - \mathbf{D}\mathbf{r}(t)|\leq \omega\label{eq:errw}
\end{equation}
and the error goes to zero after every population spike. More accurate representations correspond to smaller scale of the thresholds and higher population firing rates of the network. 

For finite size networks, the error may exceed the value of $\omega$ (whenever $\boldsymbol{\hat d}(t)$ exits the ball $|\boldsymbol{\hat d}| = \omega$). The exact bound on the representation error will depend on the set of vectors of the feedforward weights $\mathbf{F_i}$. 
Also, the error right after the spike will not be exactly zero, but will be bounded in norm by $\Delta$, which is the largest distance from the center of a firing face to an adjacent corner (see (\ref{eq:Delta})). 

When the firing surface is open (see figures \ref{fig:box}b,c), the error grows unboundedly when the input is in the certain sector of the space. 
The minimal number of neurons needed to avoid the situation of figure \ref{fig:box}b for a $J$-dimensional input is $J+1$, in which case the state domain $\cal{D}$ is a J-dimensional symplex with J+1 faces (triangle, tetrahedron, and so on). However, in this case the angles between neighboring firing faces are less than 90 degrees, which implies that the state vector can be ``bounced'' outside of the state domain and it is not possible to guarantee that only one neuron at a time is above threshold, which is our assumption. The smallest number of neurons that allows to avoid this situation is $2J$, in which case the feedforward weights $\boldsymbol{F}$ should be chosen in such a way, that the firing surface is a hypercube. 
 
The representation is most accurate when the sizes of the firing faces are the smallest. As there is no a priori reason to single out any direction in the state space, we require the size of the largest firing face $\Delta$ to be as small as possible, which, given final number of neurons $N$, means distributing the vectors $\boldsymbol{F_i}$ approximately uniformly on the $J-1$ dimensional sphere. 
 
The apparent contradiction pointed at in the beginning of this section, namely that the fact of a single voltage reaching the threshold value (a spike of a particular neuron) unambiguously determines the instantaneous value of the $J$-dimensional state vector, is resolved as follows: the spike of a given neuron $i$ implies not only that the voltage of this neuron has reached the threshold, but also, that the voltage of this neuron has reached the threshold first, before the voltage of any other neuron in the network did.  

\section{Slow connections. Voltage-to-input predictability}\label{sec:SB}
In this section we consider a network with additional synaptic currents that are slow on the time scale of neuronal integration $\frac{1}{\lambda}$. Now the dynamics of the network is described by the following equation
\begin{align}
&\dot V_i(t) = -\lambda V_i(t) + \sum_{\alpha = 1}^JF_{i\alpha}c_\alpha(t) + \sum_{j = 1}^N\Omega^\text{f}_{ij}\sum_{s = 1}^{N_j}\delta(t-t_s^j)+ \sum_{j = 1}^N\Omega^\text{s}_{ij}h^\text{s}_j(t) \hspace{0.5cm} \text{for  } i = 1\dots N\notag \\
&V_i(t_s^i) = T_i\hspace{0.5cm} \text{spikie of neuron  }i 
\end{align}
The slow synaptic current has a kernel that decays exponentially with the decay constant $\lambda_s\ll \lambda$

\begin{equation}
h^\text{s}_i(t) = \sum_{s = 1}^{N_i}\text{e}^{-\lambda_s(t-t_s^i)}\label{eq:slowH}
\end{equation}

We assume the connectivity matrix, corresponding to the slow synapses, to be of the same form as the fast connectivity matrix, but with a different decoder $\boldsymbol{D^\text{s}}$:
\begin{equation}
\Omega^\text{s}_{ij} = - \sum_{\alpha = 1}^JF_{i\alpha}D^\text{s}_{j \alpha}
\label{eq:OmegaSB}
\end{equation}

It is also crucial for what is described in this section, that the input changes slowly on the time scale of neuronal time constant $\frac{1}{\lambda}$. 

The equation (\ref{eq:evdhat}) for time evolution of the state vector $\boldsymbol{\hat d}$ becomes: 
\begin{align}
&\dot {\hat d}_\alpha(t) = -\lambda\hat d_{\alpha}(t) + c_\alpha(t) - \sum_{j = 1}^ND^\text{f}_{j \alpha}\sum_{s = 1}^{N_j}\delta(t-t_s^j) - \sum_{j = 1}^ND^\text{s}_{j \alpha}\sum_{s = 1}^{N_j} \text{e}^{-\lambda_s(t-t_s^j)} 
\end{align}
which can be solved between the spikes, assuming that the previous spike happened at the moment $t_0$
\begin{equation}
\hat d_\alpha(t) = \int_{t_0}^t\left(c_\alpha(t') - \sum_{j = 1}^ND^s_{j \alpha}h^\text{s}_j(t')\right)\text{e}^{-\lambda(t-t')}dt' + \hat d_\alpha(t_0)\text{e}^{-\lambda(t-t_0)}
\label{eq:ibs}
\end{equation} 
The next spike will happen when the vector $\boldsymbol{\hat d}(t)$ will reach the firing face of some neuron $i$: $\boldsymbol{\hat d}(t_s^i) = \boldsymbol{\hat d^{(i)}}$, where $\boldsymbol{\hat d^{(i)}}$ is given by (\ref{eq:di}). If this happens at the time $t_s^i$ such that $\lambda(t_s^i-t_0)\gg 1$, we can ignore the last term in (\ref{eq:ibs}).
Taking into account that $\boldsymbol{c}(t)$ and $\boldsymbol{h^\text{s}}(t)$ are changing slowly, we can write:
\begin{equation}
\hat d_\alpha (t_s^i) = \hat d^{(i)}_\alpha \approx \frac{1}{\lambda}\left(c_\alpha(t_s^i) - \sum_{j = 1}^ND^s_{j \alpha}h^\text{s}_j(t_s^i)\right) 
\label{eq:dI}
\end{equation}
This implies, that the identity of the spiking neuron determines approximately the total input, feedforward plus recurrent, at the time of the spike. In the limit of infinitely large $\lambda$, the equation is exact. 
Thus, the network "knows" what correction is needed to be made by the spiking neuron in order to bring the total input into all the neurons to zero, not only the current voltages as in the previous section. 

As at the moment of the spike of neuron $i$, the $i$-th component of the slow synaptic current $h_i^\text{s}$ is increased by 1, it follows from (\ref{eq:dI}) that to bring the total current to zero after the spike, the corresponding column of the matrix $\boldsymbol{D^\text{s}}$ should be chosen as follows
\begin{equation}
\mathbf{D^{s}_{i}} = \lambda\mathbf{\hat d^{(i)}} = \lambda\omega\frac{\mathbf{F_i}}{|\mathbf{F_i}|}
\label{eq:DSB}
\end{equation}

As the recurrent input decays with the slow time-constant set by the synaptic dynamics, and the feedforward input evolves also slowly but with its own dynamics, the two inputs will eventually diverge and stop cancelling each other. As this happens, the state vector $\mathbf{\hat d}(t)$, and thus the voltages of the neurons will slowly deviate from zero. 
In summary,
having both fast and slow types of connections in place ensures that the voltages go to zero very fast after the spike, and then evolve slowly as the recurrent input ceases to match the feedforward one. Removing the fast connections while leaving the slow ones in place, will not change the picture drastically: instead of going to zero right after a spike, the state vector $\mathbf{\hat d}(t)$ will decay to zero with the fast time scale $\lambda$, and then deviate from zero again as the recurrent inputs ceases to match the feedforward one. 

\section{Temporally-structured input. Expanding the dimensionality of the network dynamics relative to the dimensionality of the feedforward input.} \label{sec:TS}
In this section we consider a feedforward input which apart from having a low-dimensional structure $\boldsymbol{I^{ext}}(t) = \boldsymbol{Fc}(t)$ is predictable in time. In particular, the principal components of the input $c_\alpha(t)$ follow a first order differential equation:
\begin{equation}
\dot c_\alpha(t) = \sum_{\beta = 1}^JA_{\alpha\beta}c_\beta(t)\hspace{1cm}\alpha = 1\dots J
\label{eq:dynsys}
\end{equation} 
where $A$ is a $J$-by-$J$ matrix whose eigenvalues are either imaginary or have negative real parts.

In this case the current value of the feedforward input $\boldsymbol{c}(t)$ can be recovered from its leaky integral, since
\begin{equation}
\int_{-\infty}^t\boldsymbol{c}(t')\text{e}^{-\lambda(t-t')}dt' = (\lambda\boldsymbol{I}+\boldsymbol{A})^{-1}\boldsymbol{c}(t)
\end{equation}

As the dynamics of the slow recurrent input is also fixed and of the first order, the instantaneous value of the recurrent input can be recovered from its filtered version in the same way. This makes it possible, in principle, to balance the total input by the contribution of the spiking neuron exactly (in the limit of the large nework) even for finite values of $\lambda$. Moreover, as the time derivatives of both, the feedforward and the recurrent inputs are determined by their current values, the derivative of the current total input and the balancing contribution can be matched to prolong the balance after a spike and delay the next one.  

We will first describe the network that balances the feedforward input with the recurrent input, and then extend the argument to construct a network that also matches the derivatives. 

\subsubsection{Balancing the predictable input.}\label{sec:DE1}
As in the previous section, we start by assuming two types of synaptic currents, one fast and one slow. The dynamics of the network is described by
\begin{align}
&\dot V_i(t) = -\lambda V_i(t) + \sum_{\alpha = 1}^JF_{i\alpha}c_\alpha(t) + \sum_{j = 1}^N\Omega^\text{f}_{ij}\sum_{s = 1}^{N_j}\delta(t-t_s^j)+ \sum_{j = 1}^N\Omega^\text{s}_{ij}h^\text{s}_j(t) \hspace{0.5cm} \text{for  } i = 1\dots N\notag\\  
&V_i(t_s^i) = T_i\hspace{0.2cm}\hspace{0.5cm} \text{:  spike of neuron  }i 
\end{align}
with the slow current $\boldsymbol{h^\text{s}}(t)$ given by (\ref{eq:slowH}). The only difference is that we assume that $\boldsymbol{c}(t)$ follows the dynamics (\ref{eq:dynsys}).

We can integrate this equation, assuming that the last spike has occurred at the moment $t_0$:
$$
\boldsymbol{V}(t) = \boldsymbol{F}(\lambda\boldsymbol{I}+\boldsymbol{A})^{-1}\boldsymbol{c}(t) + \frac{1}{\lambda-\lambda_\text{s}}\boldsymbol{\Omega^\text{s}h^\text{s}}(t) + \boldsymbol{V_0}\text{e}^{-\lambda(t-t_0)}
$$
where $V_0$ is a constant of integration.

At the time of the next population spike, $t_s^i$,  the last term can be ignored, assuming that interspike interval is large compared to $1/\lambda$.
\begin{equation}
\boldsymbol{V}(t_s^i) = \boldsymbol{F}(\lambda\boldsymbol{I}+\boldsymbol{A})^{-1}\boldsymbol{c}(t_s^i) + \frac{1}{\lambda-\lambda_\text{s}}\boldsymbol{\Omega^\text{s}h^\text{s}}(t_s^i)
\end{equation}
The vector of total currents into all the neurons at the moment just before the spike is 
 \begin{equation}
 \boldsymbol{I^\text{tot}}(t_s^i)  = \boldsymbol{Fc}(t_s^i) + \boldsymbol{\Omega^\text{s}h^\text{s}}(t_s^i)\label{eq:iim} 
\end{equation}
 and it is this value that should be balanced by the contribution of the spike of the neuron $i$ to the slow current (which is given by the corresponding column of the matrix $\Omega^\text{s}$). In order for this to be possible, the value of expression (\ref{eq:iim}) should be the same every time neuron $i$ spikes. 
 
 If we had chosen the matrix of slow connectivity $\boldsymbol{\Omega^\text{s}}$ to be of the form $\boldsymbol{\Omega^\text{s}} = -\boldsymbol{FD^\text{s}}$ as before, it is the value of $(\lambda\boldsymbol{I} +\boldsymbol{A})^{-1}\boldsymbol{c}(t) + \frac{1}{\lambda-\lambda_\text{s}}\boldsymbol{h^\text{s,before}}$ that would be determined by the preferred vector of the neuron $i$, $\boldsymbol{\hat{d}^{(i)}}$, and thus be the same at every instance of neuron $i$ firning a spike. However, the value of this expression implies a certain value of (\ref{eq:iim}) only approximately, when we ignore $|\boldsymbol{A}|$ and $\lambda_\text{s}$ in comparison to $\lambda$, which we have done in the previous section. The norm of the matrix, in this case $|\boldsymbol{A}|$, is defined as the maximum of the absolute value of its eigenvalues. We keep this convention throughout the current manuscript.  
 
To ensure that the identity of the spiking neuron determines the value of the total input current (\ref{eq:iim}) exactly, the dimensionality of the network dynamics should be extended relative to the dimensionality of the feedforward input by assuming the matrix of slow connections to be of the form:
 
\begin{equation}
\boldsymbol{\Omega^\text{s}} = -\boldsymbol{FD^\text{s}} +\boldsymbol{ F^\text{int}\tau}\boldsymbol{D^\text{s}},\label{eq:Os2}
\end{equation}
 with the columns of the $N$ by $J$ matrix $\boldsymbol{F^\text{int}}$ being orthogonal to the  F-subspace
 $$
 \sum_{i = 1}^N F_{i\alpha}F^\text{int}_{i\beta}  = 0\hspace{0.6cm}\forall \alpha,\beta
 $$
 
 The vector of the total input currents then has two components: along the subspace $\text{Im}\boldsymbol{F}$ and along the subspace $\text{Im}\boldsymbol{F^\text{int}}$
 \begin{equation}
 \boldsymbol{I^\text{tot}}(t_s^i) = \boldsymbol{F}(\boldsymbol{c}(t_s^i)-\boldsymbol{D^\text{s}h^\text{s}}(t_s^i)) + \boldsymbol{F^\text{int}\tau D^\text{s}h^\text{s}}(t_s^i)\label{eq:iim2} 
 \end{equation}

 We assume that the $2J$-dimensional vectors obtained by concatinating the corresponding rows of the matrices $\boldsymbol{F}$ and $\boldsymbol{F^\text{int}}$,  written as vectors $\begin{bmatrix}\boldsymbol{F_i}\\ \boldsymbol{F^\text{int}_i}\end{bmatrix}$ have a particular distribution in the $2J$-dimensional space, that we will discuss later. 
 The matrix $\boldsymbol{\tau}$ is an invertible $J$ by $J$ matrix. We will also discuss the constrains on the eigenvalues of $\boldsymbol{\tau}$ later in this section. 
 
The dynamics of the network is now described by the $2J$-dimensional state vector $\boldsymbol{\hat d}(t)$, that is bounded by the firing surface, which is now embedded in the $2J$-dimensional space. The location of the firing face of each neuron is described by $2J$-dimensional vector $\boldsymbol{\hat d^{(i)}}$. The vectors $\boldsymbol{\hat d^{(i)}}$ are determined by the feedforward weights of the neurons $\boldsymbol{F_i}$ and the corresponding rows $\boldsymbol{F^\text{int}_i}$ of the matrix $\boldsymbol{F^\text{int}}$ 
\begin{equation}
\boldsymbol{\hat d^{(i)}} = \frac{T_i}{\boldsymbol{F_i^2}+ (\boldsymbol{F^\text{int}_i})^2}\begin{bmatrix}
\boldsymbol{F_i}\\
\boldsymbol{F^\text{int}_i}
\end{bmatrix} = \omega\frac{1}{\sqrt{\boldsymbol{F_i^2} + (\boldsymbol{F^\text{int}_i}})^2}\begin{bmatrix}
\boldsymbol{F_i}\\
\boldsymbol{F^\text{int}_i}
\end{bmatrix}
\label{eq:di2}
\end{equation}  

The fast interactions are now responsible for bringing the $2J-$dimensional vector $\boldsymbol{d}(t)$ to the origin, and hence the rank of the matrix
 of fast connections $\boldsymbol{\Omega^\text{f}}$ becomes $2J$, rather than $J$ as in the previous section
 
$$
\boldsymbol{\Omega^\text{f}} = - [\boldsymbol{F}\hspace{0.1cm} \boldsymbol{F^\text{int}}]\boldsymbol{D^\text{f}}
$$
where the columns of the matrix $\boldsymbol{D^\text{f}}$ are given by the selectivity vectors $\boldsymbol{\hat d^{(i)}}$ from (\ref{eq:di2}).

Now the spike of neuron $i$ means that two $J$-dimensional vector equations are satisfied:
\begin{equation}
\left\{\begin{matrix}
(\lambda\boldsymbol{I} + \boldsymbol{A)}^{-1}\boldsymbol{c}(t_s^i)-\frac{1}{\lambda-\lambda_\text{s}}\boldsymbol{D^\text{s}h^\text{s}}(t_s^i) = \boldsymbol{\hat d^{(i)}_{1\dots J}}\\
\frac{1}{\lambda-\lambda_\text{s}}\boldsymbol{\tau}\boldsymbol{D^\text{s}h^\text{s}}(t_s^i) = \boldsymbol{\hat d^{(i)}_{J+1\dots 2J}}\
 \end{matrix}
 \right.\label{eq:ASCDE1}
 \end{equation}
These two equation can be solved together to determine $\boldsymbol{c}(t_s^i)$ and $\boldsymbol{D^\text{s}h^{s}}(t_s^i)$ separately:
\begin{align}
&\boldsymbol{D^\text{s}h^\text{s}}(t_s^i) = (\lambda-\lambda_\text{s})\boldsymbol{\tau}^{-1}\boldsymbol{\hat d^{(i)}_{J+1\dots 2J}}\notag\\
&\boldsymbol{c}(t_s^i) = (\lambda\boldsymbol{I}+\boldsymbol{A})(\boldsymbol{\hat d^{(i)}_{1\dots J}}+\boldsymbol{\tau}^{-1}\boldsymbol{\hat d^{(i)}_{J+1\dots 2J}})\notag
\end{align}

This implies that the spike of
neuron $i$ specifies not only the instantaneous value of the total input filtered with the neuronal time constant (a particular linear combination of the instantaneous values of the feedforward and the recurrent inputs), but also about the instantaneous values of the feedforward and the recurrent inputs separately. In other words, every time the neuron $i$ spikes, $\boldsymbol{c}(t)$ is the same, and so is $\boldsymbol{D^\text{s}h^{s}}(t)$. This allows to correct for the total input imbalance (\ref{eq:iim}) with the slow synaptic current of the neuron $i$, or more precisely, the component of the total input along the subspace $\text{Im}\boldsymbol{F}$ (the first term in (\ref{eq:iim2})). As the contribution of the spike of the neuron $i$ to this component of the synaptic current is equal to $-\boldsymbol {FD^\text{s}_i}$, this is achieved by choosing

\begin{equation}
\boldsymbol{D^\text{s}_i} = \boldsymbol{c}(t_\text{s}^i) - \boldsymbol{D^\text{s}h^\text{s}}(t_\text{s}^i) = (\lambda\boldsymbol{I} + \boldsymbol{A})\boldsymbol{\hat d^{(i)}_{1\dots J}} + (\lambda_\text{s}\boldsymbol{I} + \boldsymbol{A})\boldsymbol{\tau}^{-1}\boldsymbol{\hat d^{(i)}_{J+1\dots 2J}}
\label{eq:DS2J}
\end{equation}
Imposing this equation for every neuron $i$ determines the matrix $\boldsymbol{D^\text{s}}$, and hence the matrix of slow connection $\Omega^\text{s}$ (see (\ref{eq:Os2})).

 In the limit of large networks, $N\rightarrow\infty$ and $J$ remaining finite, the total input current in the $F$-subspace can be balanced exactly after every spike. However, this is achieved at the cost of introducing an additional current in the subspace  $\text{Im}\boldsymbol{F^\text{int}}$, which will also drive the neurons away from the rest potential, shortening interspike interval. In order for the dimensionality expansion described above to make the network more efficient, we must require that this extra current is small in comparison with the imbalance that remains after a spike in the case of no dimensionality expansion and $\boldsymbol{D^\text{s}} = \lambda \boldsymbol{D^\text{f}x}$ (see section \ref{sec:SB}). This imbalance is of the order of $\frac{\lambda^\text{slow}}{\lambda} \boldsymbol{c}(t_s^i)\approx \frac{\lambda^\text{slow}}{\lambda}\boldsymbol{D^\text{s}h^\text{s}}(t_s^i)$, where 
 $$
\lambda^\text{slow} = \text{max}(|\boldsymbol{A}|,\lambda^\text{s})
$$ 
 is the slowest time scale in the problem. So, in order for the newly introduced second term of equation (\ref{eq:iim2}) to be small in comparison with the first term, we require that $|\boldsymbol{\tau}|\boldsymbol{D^\text{s}h^\text{s}}(t_s^i)\ll \frac{\lambda^\text{slow}}{\lambda}\boldsymbol{c}(t_s^i)$, or
 \begin{equation}
 |\boldsymbol{\tau}|\ll \frac{\lambda^\text{slow}}{\lambda}\label{eq:tauSmall}
 \end{equation}
 
This condition alone would imply that it is best to choose the norm of the matrix $\boldsymbol{\tau}$ as small as possible to maximize the representation efficiency. However, this is only true in the limit of $N\rightarrow\infty$. For finite networks, the condition (\ref{eq:ASCDE1}) is guaranteed to be satisfied after every spike only with some accuracy determined by the bound on the size of firing faces of neurons, $\Delta$ (see section \ref{sec:fastOnly}). The final size effect introduces the error of the order $\Delta$ into the the second equation of (\ref{eq:ASCDE1}), which translates into a residual imbalance in the total synaptic current after the spike, whose upper bound inversely scales with the norm of the matrix $\boldsymbol{\tau}$ and is equal to $(\lambda_\text{s}\boldsymbol{I} +\boldsymbol{A})\boldsymbol{\tau}^{-1}\boldsymbol{\Delta}(t_s^i)$. This imbalance should be small compared to the correction to the current made by the spike, which is of norm $\lambda\omega$. Consequently, there is another constraint on the norm of the matrix $\tau$:

\begin{equation}
|\boldsymbol{\tau}|\gg\frac{\lambda^\text{slow}}{\lambda}\frac{\Delta}{\omega}
\label{eq:tauBig}
\end{equation}   
Here $\frac{\Delta}{\omega}$ is the maximum value of the angle at which a firing face is seen from the origin. This value depends on the number of neurons in the network and the distribution of the vectors $\boldsymbol{\hat d^{(i)}}$, but not on the representation error $\omega$, which uniformly scales the firing domain but does not change the angles. 

In the network of section \ref{sec:SB}, whose dynamics has the same dimensionality as its input, it was most efficient to assume approximately uniform distribution of the directions of the firing vectors of the neurons $\boldsymbol{\hat d_i}$ in the $J$-dimensional space. In the present case, however only a small fraction of the directions in the $2J-$dimensional state space will be explored by the network. This implies that the number of neurons necessary to ensure that the firing faces are small enough, is much lower than would be expected if the entire sphere in the $2J-$ dimensional space had to be tiled. In the Appendix, section \ref{sec:App}, we describe the most efficient arrangement of the firing vectors $\boldsymbol{\hat d_i}$ (which are collinear with the vectors $[\boldsymbol F, \boldsymbol F^\text{int}]$, obtained by concatenating the corresponding rows of the matrices $\boldsymbol{F}$ and $\boldsymbol{F^\text{int}}$), and derive from the condition (\ref{eq:tauBig}) an approximate lower bound on the number of neurons in the network $N$:

\begin{equation}
N\gg \frac{(2B)^JV_JS_J}{V_{2J-1}\omega^J|\boldsymbol \tau|^{J-1}}\left(\frac{\lambda^\text{slow}}{\lambda}\right)^{2J-1}
\label{eq:boundN}
\end{equation}
where $S_J$ is the area of the unit sphere and $V_J$ is the volume of the unit ball in $J-$ dimensional space, $B$ is an upper bound on the typical value of $|\boldsymbol{c}|/\lambda$.

 If this lower bound is not satisfied, the proposed scheme of expanding the dimensionality of the network dynamics in comparison with the dimensionality of the input will not make the input representation more efficient. 

\subsubsection{Balancing the predictable input with matching the first derivative}\label{sec:DE2}
In this section we consider a network with two types of slow synaptic currents, $\boldsymbol{h^{\text{s1}}}$ and $\boldsymbol{h^{\text{s2}}}$ that decay with different rates $\lambda_{\text{s1}}$ and $\lambda_{\text{s2}}$ 
\begin{align}
\dot{h}^{\text{s1}}_i(t) = -\lambda_{\text{s1}}h^{\text{s1}}_i(t) + \rho_i(t)\notag\\
\dot{h}^{\text{s2}}_i(t) = -\lambda_{\text{s2}}h^{\text{s2}}_i(t) + \rho_i(t)
\end{align}
where $\rho_i(t)$ is neural response function of the neuron $i$ (the train of Dirac delta-functions at spike times).

We will show that for feedforward input following (\ref{eq:dynsys}), we can choose the recurrent connectivity in such a way that not only the values of the recurrent and the feedforward inputs are guaranteed to match after every spike, but also their first derivatives. This prolongs the feedforward-recurrent balance after a spike and increases the interspike interval further, making the representation of the input even more efficient. 

The network now has two slow connectivity matrices $\boldsymbol{\Omega^{\text{s1}}}$ and $\boldsymbol{\Omega^\text{s2}}$ associated with the two currents. The expression for the voltage between the spikes becomes  
\begin{equation}
\boldsymbol{V}(t) = \boldsymbol{F}(\lambda\boldsymbol{I}+\boldsymbol{A})^{-1}\boldsymbol{c}(t) + \frac{1}{\lambda - \lambda_\text{s1}}\boldsymbol{\Omega^\text{s1}h^\text{s1}}(t) + \frac{1}{\lambda - \lambda_\text{s2}}\boldsymbol{\Omega^\text{s2}h^\text{s2}}(t) + \boldsymbol{V_0}\text{e}^{-\lambda(t-t_\text{sp})}
\label{eq:V52}
\end{equation}
where $t_\text{sp}$ is the time of the last population spike, and $\boldsymbol{V_0}$ is the integration constant that is determined from the value of the voltage right after the spike (if the matrix of fast connections is in place, $\boldsymbol{V_0}$ is determined from $\boldsymbol V(t) =\boldsymbol 0$).

The dimensionality of the network dynamics is expanded even further compared to the dimensionality of the input, and is now equal to $3J$. 
The slow connectivity matrices are of the form:
\begin{align}
\boldsymbol{\Omega^\text{s1}} = -\boldsymbol{FD^\text{s1}} + \boldsymbol{F^\text{int}\tau_1D^\text{s1}} +\boldsymbol{\bar F^\text{int}\bar\tau_1D^\text{s1}} \notag\\
\boldsymbol{\Omega^\text{s2}} = -\boldsymbol{FD^\text{s2}} + \boldsymbol{F^\text{int}\tau_2D^\text{s2}} +\boldsymbol{\bar F^\text{int}\bar\tau_2D^\text{s2}} 
\end{align}
Here, $\boldsymbol{F^\text{int}}$ is an $N\times J$-dimensional matrix whose columns are orthogonal to the columns of the matrix $\boldsymbol{F}$, and $\boldsymbol{\bar F^\text{int}}$ is another $N\times J$ matrix with the columns orthogonal to all the columns of matrices $\boldsymbol{F}$ and $\boldsymbol{F^\text{int}}$, and $\boldsymbol{\tau_1}$, $\boldsymbol{\tau_2}$, $\boldsymbol{\bar\tau_1}$ and $\boldsymbol{\bar\tau_2}$ are invertible $J\times J$ matrices that are arbitrary except for the scale of their eigenvalues. 

The state vector of the network $\boldsymbol{\hat d}(t)$ is now $3J$ - dimensional and so are the firing surface and the selectivity vectors of neurons, $\boldsymbol{\hat d^{(i)}}$. 

The neuron $i$ spikes when the state vector of the network $\boldsymbol{\hat d}(t)$ reaches the corresponding firing face. As usual we assume, that the number of neurons $N$ is large, the vectors $[\boldsymbol {F,F^\text{int},\bar F^\text{int}} ]$ cover uniformly the directions of the $3J$-dimensional space that are explored by the state vector $\boldsymbol{\hat d}(t)$, and that the interspike interval is large on the time scale of $1/\lambda$ so that the last term in equation (\ref{eq:V52}) can be ignored at the moment right before a spike. Then, this implies that three vector equation are satisfied at the moment just before the spike:

\begin{equation}
\left\{
\begin{matrix}
(\lambda\boldsymbol{I} + \boldsymbol{A})^{-1}\boldsymbol{c}(t_\text{s}^i) - \frac{1}{\lambda-\lambda_\text{s1}}\boldsymbol{D^\text{s1}h^\text{s1}}(t_\text{s}^i) - \frac{1}{\lambda-\lambda_\text{s2}}\boldsymbol{D^\text{s2}h^\text{s2}}(t_\text{s}^i)= \boldsymbol{\hat d^{(i)}_{1\dots J}}  \\
\frac{1}{\lambda-\lambda_\text{s1}}\boldsymbol{\tau_1D^\text{s1}h^\text{s1}}(t_\text{s}^i) +  \frac{1}{\lambda-\lambda_\text{s2}}\boldsymbol{\tau_2D^\text{s2}h^\text{s2}}(t_\text{s}^i) = \boldsymbol{\hat d^{(i)}_{J+1\dots 2J}}\\

\frac{1}{\lambda-\lambda_\text{s1}}\boldsymbol{\bar\tau_1D^\text{s1}h^\text{s1}}(t_\text{s}^i)+  \frac{1}{\lambda-\lambda_\text{s2}}\boldsymbol{\bar\tau_2D^\text{s2}h^\text{s2}}(t_\text{s}^i)= \boldsymbol{\hat d^{(i)}_{2J+1\dots 3J}}
\end{matrix} 
\right.
\label{eq:SCDE2}
\end{equation}
The balance condition is imposed only in the input subspace (the subspace spanned by the columns of the matrix $\boldsymbol{F}$), and the recurrent input in the additional dimensions is small. We require that both, the total input and its first derivative are zero after a spike:
\begin{align}
&\boldsymbol{c}(t_\text{s}^i) - \boldsymbol{D^\text{s1}h^\text{s1,after}}(t_\text{s}^i) - \boldsymbol{D^\text{s1}h^\text{s1,after}}(t_\text{s}^i) = \mathbf{0}\notag\\
&\boldsymbol{Ac}(t_\text{s}^i) +\lambda_\text{s1} \boldsymbol{D^\text{s1}h^\text{s1,after}}(t_\text{s}^i) +\lambda_\text{s2}\boldsymbol{D^\text{s1}h^\text{s1,after}}(t_\text{s}^i) = \mathbf{0}
\label{eq:balanceDE2}
\end{align}

This implies that the contributions of spiking neuron $i$ to the two synaptic currents along the input space, $\boldsymbol{D^\text{s1}_i}$ and $\boldsymbol{D^\text{s2}_i}$, should satisfy
\begin{align}
&\boldsymbol{D^\text{s1}_i} + \boldsymbol{D^\text{s2}_i} = \boldsymbol{c}(t_\text{s}^i) - \boldsymbol{D^\text{s1}h^\text{s1}}(t_\text{s}^i) - \boldsymbol{D^\text{s2}h^\text{s2}}(t_\text{s}^i)\notag\\
&\lambda_\text{s1}\boldsymbol{D^\text{s1}_i} + \lambda_\text{s2}\boldsymbol{D^\text{s2}_i} =  - (\boldsymbol{Ac}(t_\text{s}^i) +\lambda_\text{s1}\boldsymbol{D^\text{s1}h^\text{s1}}(t_\text{s}^i) +\lambda_\text{s2} \boldsymbol{D^\text{s2 }h^\text{s2}}(t_\text{s}^i)) 
\label{eq:balanceDE2i}
\end{align}

This is possible only if the time-dependent right-hand side of the equations is the same every time neuron $i$ spikes. And indeed, the system of equations (\ref{eq:SCDE2}) can be solved to obtain the components of the recurrent and feedforward currents along the input space at the moment just before the spike $\boldsymbol{D^\text{s1}h^\text{s1}}(t_\text{s}^i)$ , $\boldsymbol{D^\text{s2}h^\text{s2}}(t_\text{s}^i)$ and $\boldsymbol{c}(t_\text{s}^i)$. The result can then be plugged in into (\ref{eq:balanceDE2i}) to obtain the expressions for the $i$-th columns of the matrices $\boldsymbol{D^\text{s1}}$ and $\boldsymbol{D^\text{s2}}$, which determine the contributions of the spike of neuron $i$ to the two slow currents $\boldsymbol{h^\text{s1}}$ and $\boldsymbol{h^\text{s2}}$. 
\begin{align}
\boldsymbol{D^\text{s1}_i} = \frac{1}{\lambda_\text{s2} - \lambda_\text{s1}}\left[(\lambda_\text{s2}\boldsymbol{I}+\boldsymbol{A})(\lambda\boldsymbol{I}+\boldsymbol{A})(\boldsymbol{ \hat d^{(i)}_{1\dots J}}+\boldsymbol{b}) + (\lambda_\text{s1}\boldsymbol{I}+\boldsymbol{A})((\lambda+\lambda_\text{s2}-\lambda_\text{s1})\boldsymbol{I}+\boldsymbol{A})\boldsymbol{c}\right]    \notag\\
\boldsymbol{D^\text{s2}_i} = \frac{1}{\lambda_\text{s1} - \lambda_\text{s2}}\left[(\lambda_\text{s1}\boldsymbol{I}+\boldsymbol{A})(\lambda\boldsymbol{I}+\boldsymbol{A})(\boldsymbol{ \hat d^{(i)}_{1\dots J}}+\boldsymbol{c}) + (\lambda_\text{s2}\boldsymbol{I}+\boldsymbol{A})((\lambda+\lambda_\text{s1}-\lambda_\text{s2})\boldsymbol{I}+\boldsymbol{A})\boldsymbol{b}\right]  
\end{align}

Here $\boldsymbol{b}$ and $\boldsymbol{c}$ are the following combinations of the $J$-dimensional projections of the firing vector $\boldsymbol{\hat d^{(i)}}$ onto coordinate planes $\boldsymbol{\hat d^{(i)}_{J+1\dots 2J}}$ and $\boldsymbol{\hat d^{(i)}_{2J+1\dots 3J}}$:
\begin{align}
&\boldsymbol{b} = (\boldsymbol{\tau_1}^{-1}\boldsymbol{\tau_2} - \boldsymbol{\bar\tau_1}^{-1}\boldsymbol{\bar\tau_2})^{-1}(\boldsymbol{\tau_1}^{-1}\boldsymbol{\hat d^{(i)}_{J+1\dots 2J}} - \boldsymbol{\bar\tau_1}^{-1}\boldsymbol{\hat d^{(i)}_{2J+1\dots 3J}})\\
&\boldsymbol{c} = (\boldsymbol{\tau_2}^{-1}\boldsymbol{\tau_1} - \boldsymbol{\bar\tau_2}^{-1}\boldsymbol{\bar\tau_1})^{-1}(\boldsymbol{\tau_2}^{-1}\boldsymbol{\hat d^{(i)}_{J+1\dots 2J}} - \boldsymbol{\bar\tau_2}^{-1}\boldsymbol{\hat d^{(i)}_{2J+1\dots 3J}})
\end{align}

The firing vector $\boldsymbol{\hat d^{(i)}}$ is in turn determined by the $i$-th row of the matrices $\boldsymbol{F}$, $\boldsymbol{F^\text{int}}$ and $\boldsymbol{\bar F^\text{int}}$:
\begin{align}
&\boldsymbol{\hat d^{(i)}_{1\dots J}} = \frac{T_i}{\boldsymbol{F_i}^2 + (\boldsymbol{F^\text{int}})^2 + (\boldsymbol{\bar F^\text{int}})^2}\boldsymbol{F_i}\notag\\
&\boldsymbol{\hat d^{(i)}_{J+1\dots 2J}} = \frac{T_i}{\boldsymbol{F_i}^2 + (\boldsymbol{F^\text{int}})^2 + (\boldsymbol{\bar F^\text{int}})^2}\boldsymbol{F^\text{int}_i}\notag\\
&\boldsymbol{\hat d^{(i)}_{2J+1\dots 3J}} = \frac{T_i}{\boldsymbol{F_i}^2 + (\boldsymbol{F^\text{int}})^2 + (\boldsymbol{\bar F^\text{int}})^2}\boldsymbol{\bar F^\text{int}_i}
\end{align}

The matrix of fast connections has rank $3J$ and is given by:
$$
\boldsymbol{\Omega^\text{f}} = - [\boldsymbol{F} \hspace{0.1cm}\boldsymbol{F^\text{int}} \hspace{0.1cm}\boldsymbol{\bar{F}^\text{int}}]\boldsymbol{D^\text{f}}
$$
with the matrix $\boldsymbol{D^\text{f}}$ consisting of the vectors $\boldsymbol{\hat d^{(i)}}$.

We estimate the number of neurons in the network $N$ required for the current scheme of dimensionality expansion to be  

\begin{equation}
N\gg \frac{(2B)^{2J}V_{2J}S_J}{V_{3J-1}\omega^{2J}\tau_\text{max}^{J-1}}\left(\frac{\lambda^\text{slow}}{\lambda}\right)^{3J-1}
\label{eq:boundN3}
\end{equation}
with $S_J$ is the area of the unit sphere and $V_J$ is the volume of the unit ball in $J-$ dimensional space, $B$ is an upper bound on the typical value of $|\boldsymbol{c}|/\lambda$ and $\tau_\text{max}$ - the maximum among the eigenvalues of the matrices $\boldsymbol{\tau_1}$, $\boldsymbol{\tau_2}$, $\boldsymbol{\bar\tau_1}$ and $\boldsymbol{\bar \tau_2}$ in absolute value (see Appendix, section \ref{sec:App}).


\section{Numerical Simulations}

\begin{figure}
\includegraphics[width = 15cm]{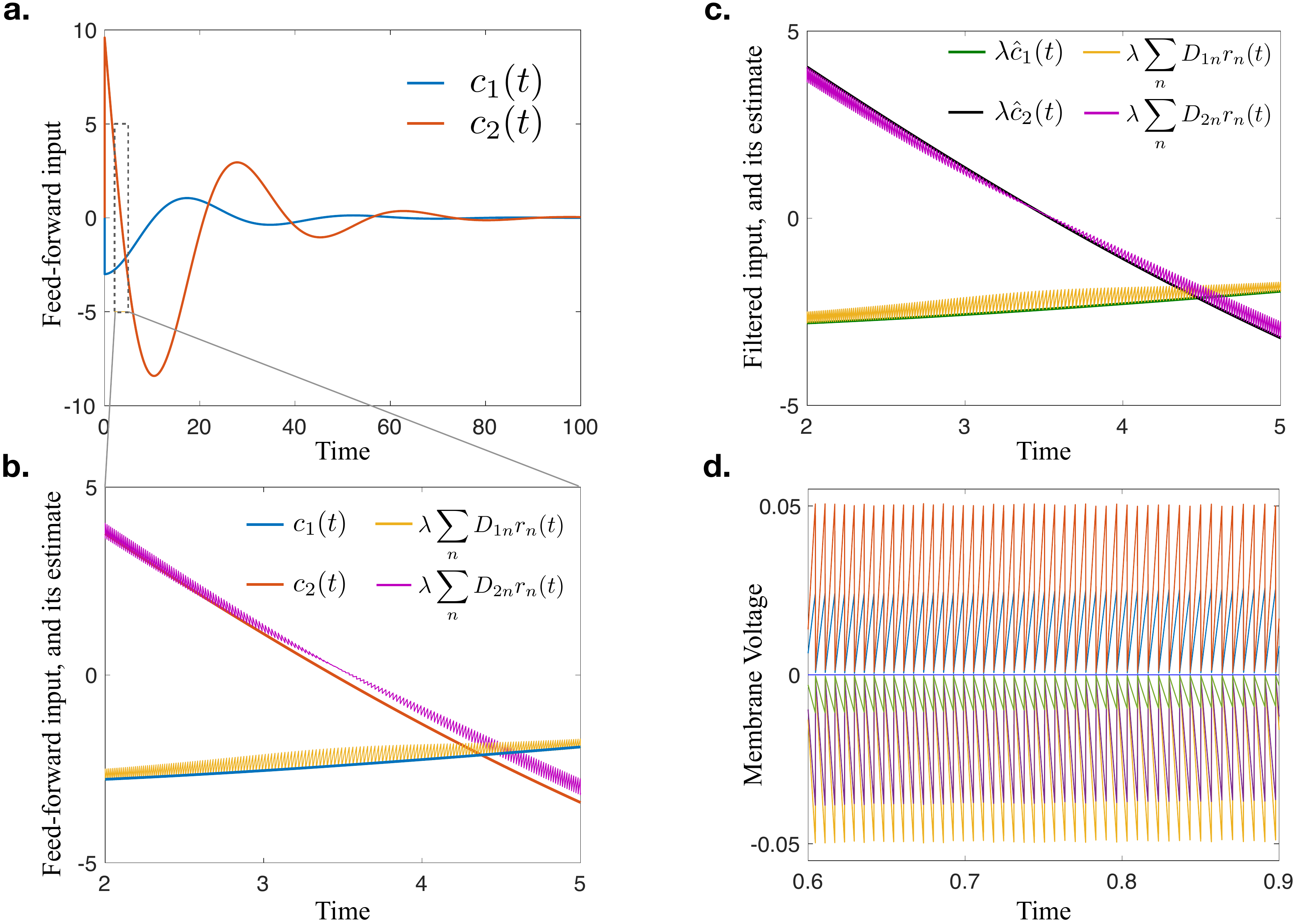}
\centering
\caption{Numerical Simulations. The network with only fast connections. \textbf{a.} The shape of two-component time varying input used in all the simulations. The total duration of the input is 100 time units, but for illustration purposes when showing the reconstructed signal we use the 3 time units window bounded by the dashed line. \textbf{b.} The two components of the input $\boldsymbol{c}(t)$ and their reconstruction from the spikes of the network $\lambda \boldsymbol{D^\text{f}r}(t)$. The reconstructed input changes stepwise at the moment of a spike approaching the true value, however not attaining it. The scale of the representational error right after a spike is determined by the ratio of the rate of change of the input and the neuronal time constant $\lambda$, which in the current case is small ($\approx 0.02$). \textbf{c.} The leaky integral of the input $\boldsymbol{\hat c}(t)$ (as defined in (\ref{eq:chat})) multiplied by $\lambda$ to be on the same scale as $\boldsymbol{c}(t)$ and the same reconstruction as in \textbf{b}, $\lambda \boldsymbol{D^\text{f}r}(t)$. The reconstruction becomes exact after every spike because the simulated neurons were chosen to have the exact required selectivity vectors (feedforward weights), imitating the infinite network which has neurons with any given selectivity vector. \textbf{d.} The trace of membrane voltages of five randomly chosen neurons. After every spike the voltages of all the neurons drop to zero in step-wise manner.}
\label{fig:sim1}
\end{figure}

To illustrate the dynamics of the efficient balanced network with different structures of synaptic interactions, we numerically simulate the activity of four different networks in response to the same feedforward input. The feedforward input in our simulation is two-dimensional and its principal components follow an autonomous linear differential equation of the first order, like in section \ref{sec:TS}: 
$
\dot{\boldsymbol{c}} = \boldsymbol{Ac}
$
with $\boldsymbol{A} = \begin{bmatrix}-0.12& -0.036\\ 1 &0\end{bmatrix}$ (eigenvalues $-0.06\pm 0.18i$) and the initial condition $\boldsymbol{c}(0) = \begin{bmatrix}-0.3\\0.96\end{bmatrix}$ (see Figure \ref{fig:sim1}\textbf{a} ).

The first network is the network with only fast synaptic currents which was described in section \ref{sec:fastOnly}. The strength of synaptic connections given by (\ref{eq:flr}) and (\ref{eq:fastD}) guarantees that the membrane voltages of all the neurons are brought to zero after every spike, which can be seen from the Figure \ref{fig:sim1}\textbf{d}, where the time evolution of the voltages of 5 randomly chosen neurons are plotted. However, this network has no capacity to balance the feedforward input between the spikes, which is reflected in the rapid rise of the membrane voltages between the spikes (see Figure \ref{fig:sim1}\textbf{d}). The total number of spikes fired by this network $N_\text{sp} = 2875$. The time evolution of the two principal components of the feedforward input $c_1(t)$ and $c_2(t)$ can be approximately recovered from the spikes of the network by means of the decoder matrix $\boldsymbol{D^\text{f}}$ given in (\ref{eq:fastD}): $\boldsymbol{c} \approx \lambda\boldsymbol{D^\text{f}r}(t)$. This reconstruction is illustrated on the Figure \ref{fig:sim1}\textbf{b}. The precision of the approximation is determined by how fast the input $\boldsymbol{c}(t)$ changes on the time scale of membrane time constant $\lambda$. In the present simulation $\lambda = 10$ and the absolute value of the eigenvalues of the matrix $A$ is $|\boldsymbol{A}|\approx 0.19\ll \lambda$. The signal that is recovered exactly in the case of an infinitely large network is the leaky integral of the feedforward input $\boldsymbol{\hat{c}}(t)$ given in (\ref{eq:chat}). The reconstruction of $\boldsymbol{\hat{c}}(t)$ as $\boldsymbol{D^\text{f}r}(t)$ is shown on the Figure \ref{fig:sim1}\textbf{c}, the plots are multiplied by $\lambda$ to keep the same scale as on the Figures \ref{fig:sim1}\textbf{a} and \ref{fig:sim1}\textbf{b}. As the feedforward weights of the neurons were chosen to correspond to the selectivity vectors required to represent this particular input (see the end of the current section for details), the reconstruction is very precise. 

\begin{figure}
\includegraphics[width = 15cm]{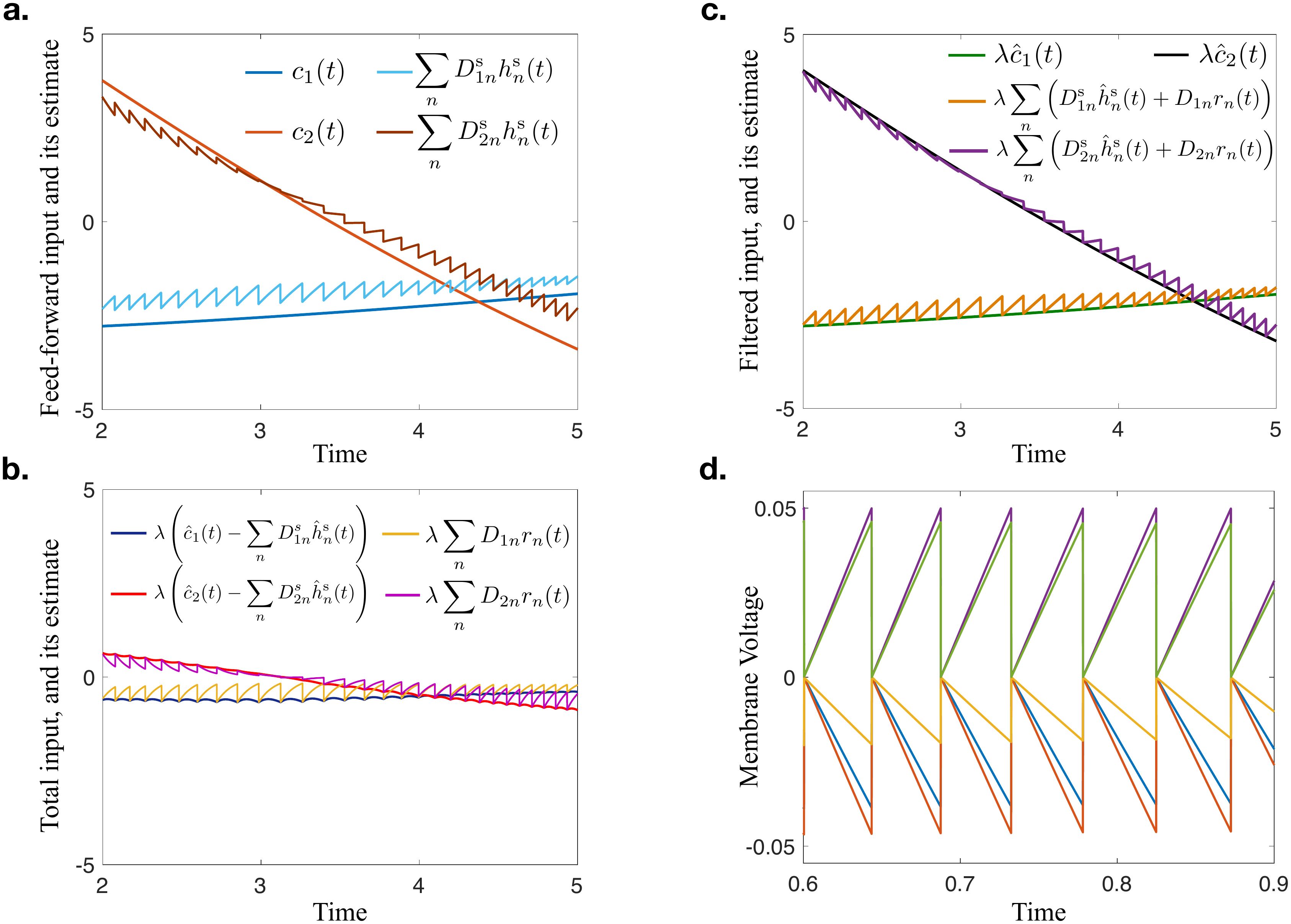}
\centering
\caption{Numerical Simulations. The network with one type of slow connections and no dimensionality expansion. \textbf{a.} The two components of the feedforward input $\boldsymbol{c}(t)$ and its reconstruction $\boldsymbol{D^\text{s}h^\text{s}}(t)$. The reconstruction is approximate with the accuracy determined by the ratio of the synaptic time constant $\lambda^\text{s}$ and the neuronal time constant $\lambda$. In the present case this ratio is equal to 0.2.  \textbf{b}. The two components of the leaky integral of the total input into the network, feedforward plus slow recurrent $\boldsymbol{\hat c}(t) - \boldsymbol{D^\text{s}\hat h^\text{s}}(t)$, together with its reconstruction $\boldsymbol{D^\text{f}r}(t)$ (both scaled by $\lambda$). The reconstruction is exact, as on the figure \ref{fig:sim1}\textbf{c}. However as the slow recurrent input approximately balances the feedforward, the amplitude of the total input is much smaller and consequently, the spikes are more sparse. \textbf{c.} The leaky integral of the feedforward input $\boldsymbol{\hat{c}}(t)$ can be recovered as the sum of the leaky integral of the recurrent input  $\boldsymbol{D^\text{s}\hat {h}^\text{s}}(t)$ and the spikes of the network filtered with neuronal time constant, $\boldsymbol{D^\text{f}r}(t)$. \textbf{d.} The evolution of membrane voltages of five randomly chosen neurons. The voltages change much more slowly compared to figure \ref{fig:sim1}\textbf{d} as the feedforward input is partially balanced by the recurrent.
}
\label{fig:sim2} 
\end{figure}

The second network that we simulate has a slowly decaying synaptic current in addition to the fast one, with the structure of the synaptic efficacies chosen in such a way, that the recurrent input is in the same 2-dimensional subspace of the neural space as the feedforward input, see equations (\ref{eq:OmegaSB}) and (\ref{eq:DSB}) and section \ref{sec:SB}. The feedforward input $\boldsymbol{c}(t)$ can now be approximately recovered from the slow synaptic current that is balancing it: $\boldsymbol{c}(t)\approx \boldsymbol{D^\text{s}h^\text{s}}(t)$. The slow current $\boldsymbol{h^\text{s}}(t)$ is in tern recoverable from the spiking activity of the network by convolving it with the decaying exponential kernel (in the present simulation, the exponent is $\lambda_\text{s} = 2$). The reconstruction is shown on the Figure \ref{fig:sim2}\textbf{a}. The precision is determined by the slowest time constant among the eigenvalues of the matrix $A$ and $\lambda_\text{s}$ in relation to $\lambda$, which in this simulation is $\lambda_\text{s}/\lambda  = 0.2$. This second network is equivalent to the first one, whose feedforward input is the sum of the actual feedforward input $\boldsymbol{Fc}(t)$ and the slow recurrent input $-\boldsymbol{FD^\text{s}h^\text{s}}(t)$. The leaky integral of this total input is exactly (in the case of infinite network) recoverable from the spikes of the network with the decoder $\boldsymbol{D^\text{f}}$ (\ref{eq:fastD}) as $\boldsymbol{\hat{c}}(t)-\boldsymbol{D^\text{s}\hat h^\text{s}}(t) = \boldsymbol{D^\text{f}r}(t)$, see Figure \ref{fig:sim2}\textbf{b}. Consequently, the signal $\boldsymbol{\hat c}$ can be decoded exactly from the combination of the spike train filtered with the exponential kernel of decay rate $\lambda$, $\boldsymbol{r}(t)$, and the slow synaptic current filtered with the same kernel, $\boldsymbol{\hat h^\text{s}}$: $\boldsymbol{\hat c}(t) = \boldsymbol{D^\text{s}\hat h^\text{s}}(t)+\boldsymbol{D^\text{f}r}(t)$, as is shown on the Figure \ref{fig:sim2}\textbf{c}. Again, as the size of the network is large, $N = 1452$ (how the number of neurons in the network and the matrix of the feedforward weights were chosen, will be explained later in this section), the reconstruction is very good. Figure \ref{fig:sim2}\textbf{d} shows the time evolution of the membrane voltages of 5 randomly chosen neurons. As the feedforward input is now partially balanced by the recurrent one, the accumulation of the voltage is slower compared to the case of the network with only fast connections (see Figure \ref{fig:sim1}\textbf{d}). The total number of spikes in response to the same feedforward input is now $N_\text{sp} = 486$. 

\begin{figure}
\includegraphics[width = 15cm]{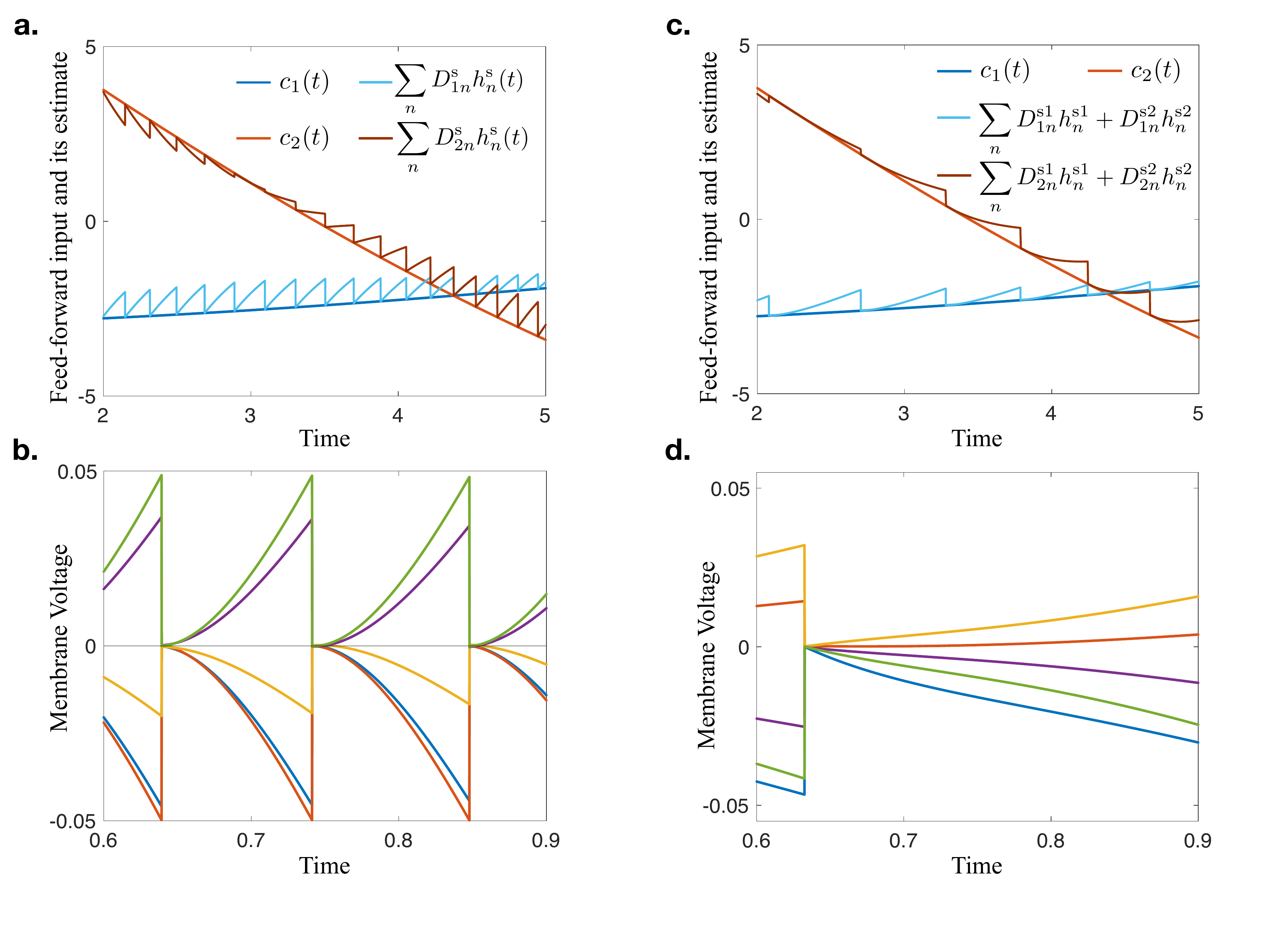}
\centering
\caption{Numerical simulations of the two networks with dimensionality expansion. \textbf{a.} The reconstruction of the input by the network with one synaptic time constant and 2-fold dimensionality expansion (section \ref{sec:DE1}). The reconstruction of the input $\boldsymbol{c}(t)$ from the slow synaptic current $\boldsymbol{D^\text{s}h^\text{s}}(t)$ is precise after every spike of the network. \textbf{b.} The membrane voltages of five randomly chosen neurons in the network of (\textbf{a}). As the feedforward input is balanced by the recurrent in the input subspace and the additional current in the $J$-dimensional internal subspace is small, the first derivatives of all the voltages are close to zero after every spike. \textbf{c.} The reconstruction of the input by the network with two synaptic time constants and 3-fold dimensionality expansion, described in the section \ref{sec:DE2}. In this case, not only the value but also the first derivative of the predictable feedforward input is matched by the total slow synaptic current after every spike. The two components of the input $\boldsymbol{c}(t)$ are encoded by the sum of the two slow synaptic currents $\boldsymbol{D^\text{s1}h^\text{s1}}(t)+\boldsymbol{D^\text{s2}h^\text{s2}}(t)$. \textbf{d.} The membrane voltages of five randomly chosen neurons in the network of (\textbf{c}). The feedforward current in the input subspace is still balanced by the synaptic current after every spike, while the additional current in the $2J$-dimensional internal subspace is larger than in (\textbf{b}), so the first derivatives of the voltages are further from zero. However, because the first derivative of the balancing synaptic current in the input subspace matches the derivative of the feedforward current, the voltages stay below threshold longer, delaying the next spike.
}
\label{fig:sim3}
\end{figure}

The results of simulating the two network structures described in the sections \ref{sec:DE1} and \ref{sec:DE2} are shown on the Figure \ref{fig:sim3}. The plots on the panel \textbf{a} of Figure \ref{fig:sim3} show the principal components of the feedforward input $c_1(t)$ and $c_2(t)$ and their reconstruction from the activity of the network of section \ref{sec:DE1}. This network still has only one synaptic time constant, $\lambda_\text{s} = 2$, but the dimensionality of its dynamics is expanded 2-fold relative to the dimensionality of the feedforward input, by not constraining the synaptic current to stay in the input subspace, see  (\ref{eq:Os2}). As  before, the feedforward input is reconstructed from the network's spikes filtered with the synaptic kernel $\text{e}^{-\lambda_\text{s}t}$, namely $\boldsymbol{h^\text{s}}$(t):  $\boldsymbol{c}(t) = \boldsymbol{D^\text{s}h^\text{s}}(t)$, where the slow decoder $\boldsymbol{D^\text{s}}$ is given by (\ref{eq:DS2J}). The error of representation is corrected exactly after every spike, which implies that the feedforward input is cancelled by the projection of the recurrent one onto the input subspace after every spike. The additional recurrent input in the subspace spanned by the columns of the matrix $\boldsymbol{F^\text{int}}$ (see (\ref{eq:Os2})) is very small, and hence the time derivative of the membrane voltages on the Figure \ref{fig:sim3}\textbf{b} are very close to zero after every spike. The total number of spikes fired by this network is $N_\text{sp} = 268$, and the number of neurons in the network is $N = 1869$.

The panels \textbf{c} and \textbf{d} of the Figure \ref{fig:sim3} refer to the network described in the section \ref{sec:DE2}, which has two slow synaptic currents decaying with time constants $\lambda_\text{s1} = 2$ and $\lambda_\text{s2} = 1.2$. The dimensionality of the network dynamics is expanded 3-fold in comparison with the dimensionality of the feedforward input.  For this network, as one can see from Figure \ref{fig:sim3}\textbf{c}, not only the feedforward input is matched by the corresponding projection of the recurrent one after every spike, but also its first time derivative, so the balance is maintained longer into the inter-spike interval. On the Figure \ref{fig:sim3}\textbf{d} we plot the membrane voltages of 5 randomly chosen neurons. As the projections of the recurrent input onto the subspaces, spanned by the columns of $\boldsymbol{F^\text{int}}$ and $\boldsymbol{\bar F^\text{int}}$ are not cancelled, the time derivatives of the voltages immediately after a spike are not zero. In fact, the derivatives right after the spike are larger than in the case of the network with only 2-fold dimensionality expansion (see Figure \ref{fig:sim3}\textbf{b}), however they stay small for a longer time, delaying the next spike of the network. This is because the main contribution to the membrane voltage, namely the feedforward input, is predicted by the network and cancelled more efficiently. 

\begin{figure}
\includegraphics[width = 15cm]{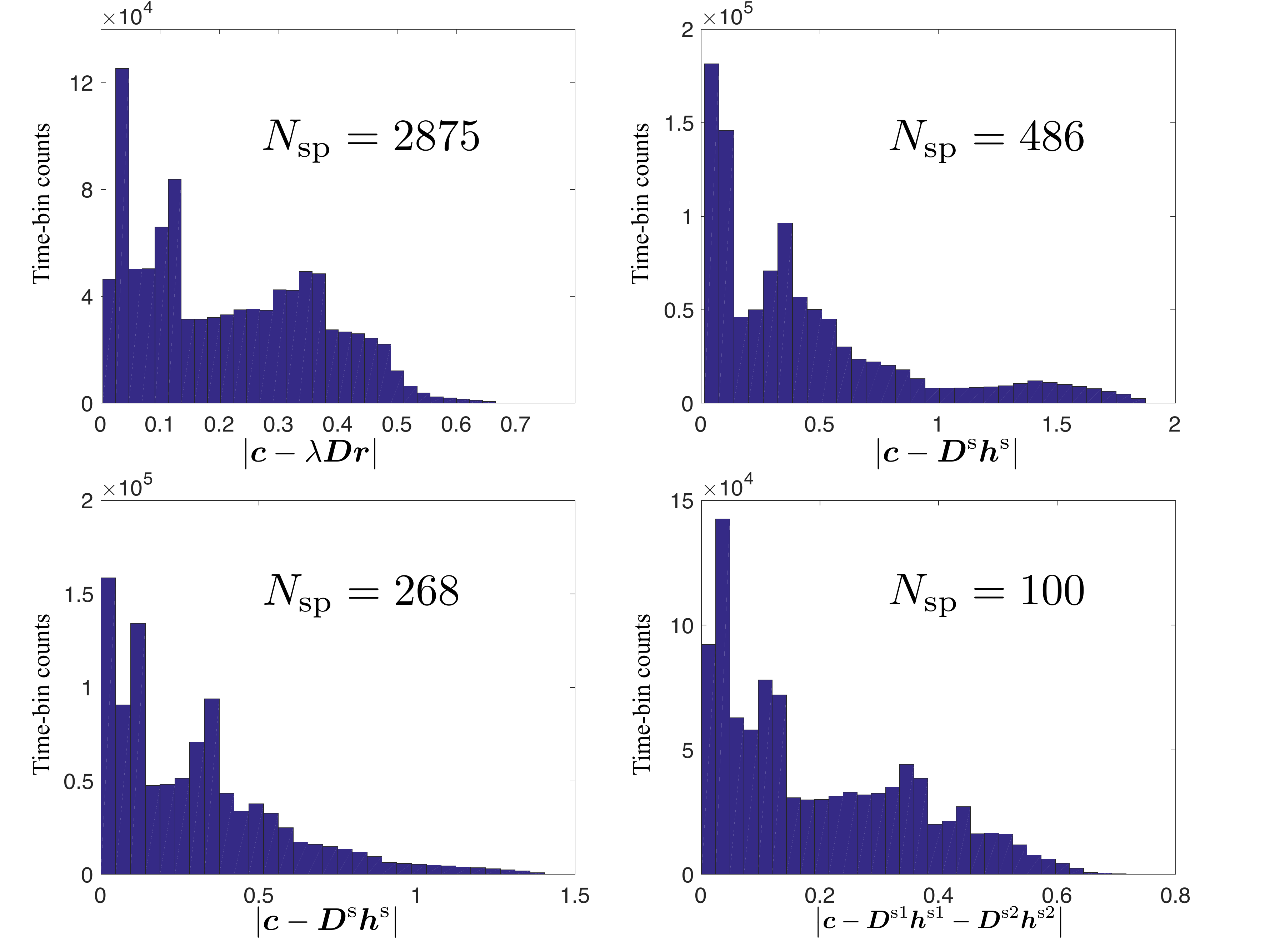}
\centering
\caption{The histograms of the absolute values of the decoding errors and the total number of spikes of the four versions of balanced network. The errors were measured at every time step of numerical simulation. \textbf{a.} The network with only fast connections. \textbf{b.} The network with one type of slow connections (one synaptic time constant) and no dimensionality expansion. \textbf{c.} The network with one type of slow connections with 2-fold dimensionality expansion. \textbf{d.} The network with 2 types of slow connections (two synaptic time constants) and 3-fold dimensionality expansion.
}
\label{fig:sim4}
\end{figure}

The increasing efficiency of the balanced network with its complexity is summarized on the Figure \ref{fig:sim4} where we plot the histogram of the absolute value of the representation error over time bins and also indicate the total number of spikes $N_\text{sp}$ fired by the corresponding network in response to the same input, which is shown on the Figure \ref{fig:sim1}\textbf{a}. 

To choose the matrix of the feedforward weights $\boldsymbol{F}$ and the matrix of the internal dimensions of the network dynamics $\boldsymbol{F^\text{int}}$ and $\boldsymbol{\bar F^\text{int}}$ we used the following procedure. We first determined the selectivity vectors $\boldsymbol{\hat d^{(i)}}$ of the neurons that would fire in response to the given input, and only generated the neurons with corresponding vectors $\boldsymbol{F_i}$, $\boldsymbol{F^\text{int}_i}$ and $\boldsymbol{\bar F^\text{int}_i}$ (if applicable). In all versions of the network except for the one with only fast connections, we also generate $2(d-1)$ (with $d$ being the dimensionality of the vector $\boldsymbol{\hat d^{(i)}}$) additional neurons per a firing neuron, whose selectivity vectors differ from $\boldsymbol{\hat d^{(i)}}$ by a small shift in a direction orthogonal to $\boldsymbol{\hat d^{(i)}}$. The additional neurons with the selectivity vectors similar to the firing neurons are there for the proof of principle.  We use this procedure to imitate the infinite network, which would have a neuron for every chosen direction in its state space but only a small number of them would be active in response to the given input. In the two cases when the dimensionality of the state space is equal to the dimensionality of the input space, namely $J = 2$ in our simulations, this was not necessary, as having very small size of the firing faces (small $\Delta$) when choosing the feedforward weights randomly would require a size of the network that is still reasonable. However, in the other two cases where the dimensionality of the state space is equal to 4 or 6, this would be impractical.

We provide the Matlab code for setting up the networks and running the simulations at \url{https://github.com/l-kushnir/temporalStr}. 

\section{Generalization of the framework}
In the setup of the previous sections, when the future evolution of the feedforward input is fully determined by its value at the current moment, one can imagine extending the dimensionality of the state space of the network and adding another slow time constant to impose the constraint that the second derivatives of the feedforward and the recurrent inputs are also matched after every spike. Similarly, the third derivative, the fourth and so forth. The representation of the input will become more and more efficient until the period into the future for which the recurrent input closely reflects the feedforward one will reach the period for which the feedforward input is predictable, i.e. follows the equation (\ref{eq:dynsys}). Matching higher order derivatives will not decrease the number of spikes further. 
It should also be noted, that as we increase the number of derivatives to be matched, the dimensionality of the firing surface increases, and the number of neurons in the network should be increased accordingly to keep the size of the firing faces small. 

\section{Cost function formulation.}
In this section we draw connection between the approach taken in this paper and the original approach of \cite{Boerlin2013}, where the network with fast connections (see section {\ref{sec:fastOnly}}) was derived as an implementation of greedy minimization of a cost function. 

We start with introducing a $J$-dimentional variable $x(t)$, that changes in time and is represented online by a sequence of discrete events. We do not specify at this point what is the nature of these events, but we refer to them as "spikes" for convenience. The online representation of the variable follows from greedy minimization of the time-dependent cost function ${\cal{L}}(t)$: 
\begin{equation}
{\cal{L}}(t) = (\boldsymbol{x}(t)-\boldsymbol{\bar x}(t))^2 + \omega^2n_\text{sp}(t)
\label{eq:cost}
\end{equation}
here $\boldsymbol{x}(t)$ is the represented variable, $\boldsymbol{\bar x}(t')$ is its estimate, $n_\text{sp}(t)$ is the number of discrete events (spikes) up to a time $t$ and $\omega$ is a parameter which specifies the cost associated with a spike.  

We assume that there are $N$ types of events that differently affect the online estimate $\boldsymbol{\bar x}(t)$. Namely, an event of type $j$ updates the estimate by a $J$-dimensional vector $\boldsymbol{D_j}$, that we refer to as the decoding vector corresponding to the event $j$. These $N$ types will later be interpreted as $N$ neurons. Between the events, the estimate decays to zero with the rate $\lambda$. 

This estimation, or decoding process can be summarized by the equation

\begin{equation}
\bar x_\alpha(t) = \sum_{j = 1}^N\sum_{s = 1}^{N_j}D_{j \alpha}\text{e}^{-\lambda(t-t_s^j)} 
\end{equation}
where $t_s^j$ is the time of the spike of a given type $j$ with the order number $s$, and $N_j$ is the total number of these spikes till the current moment $t$.

Note that 
$$
\bar x_\alpha(t) = \sum_{j = 1}^ND_{j \alpha} r_j
$$ 
where $r$ is defined in section \ref{sec:1}  (see (\ref{eq:hatr})). 

We assume that $N$ is very large and for any arbitrarily chosen vector of the representation error, there is a decoding vector $\boldsymbol{D_i}$ which is sufficiently close to it, so that any error in the representation of $\boldsymbol{x}(t)$ can be corrected at any time. However, if the current error is small this correction will not reduce the cost function ${\cal{L}}(t)$ as additional spike increases the second term in (\ref{eq:cost}) by $\omega^2$:

Since any decoding vector is available, the greedy algorithm will choose the one that eliminates the error at the time of the event. Let's say, this is the vector $\boldsymbol{D_i}$. Then
\begin{equation}
x_\alpha(t_s^i) - \bar x_\alpha(t_s^i) = D_{i \alpha}\hspace{1cm} \alpha = 1\dots J
\label{eq:eD}
\end{equation}

Greedy minimization implies that the spike reduces the value of the cost function
\begin{equation}
{\cal{L}}_\text{spike}(t_s^i) - {\cal{L}}_\text{no spike}(t_s^i) < 0.
\label{eq:dcost}
\end{equation}
If the encoded variable $\boldsymbol{x}(t)$ changes continuously, so does the ${\cal{L}}_\text{no spike}(t)$ and the spike happens at the moment when it becomes equal to ${\cal{L}}_\text{spike}(t)$:

\begin{equation}
{\cal{L}}_\text{spike}(t_s^i) - {\cal{L}}_\text{no spike}(t_s^i) = 0
\label{eq:dcost1}
\end{equation}

Together with the definition of (\ref{eq:cost}) and the equation (\ref{eq:eD}), this condition leads

\begin{equation}
-2\boldsymbol{D_i}^T(\boldsymbol{x}(t_s^i) -\boldsymbol{\bar x}(t_s^i) ) + \boldsymbol{D_i}^T\boldsymbol{D_i}+\omega^2=0
\label{eq:dcost2}
\end{equation}
Taking into account (\ref{eq:eD}), this becomes
\begin{equation}
|\boldsymbol{D_i}| = \omega,
\end{equation}
which implies that only the decoding vectors of the norm $\omega$ are relevant, and only events of the type associated to such decoding vectors will be chosen. So, we can restrict the set of decoding vectors to 
\begin{equation}
|\boldsymbol{D_i}| = \omega\hspace{1cm}i = 1\dots N
\end{equation}
Note, that this is slightly different from the line of reasoning in \cite{Boerlin2013} where the set of decoding vectors was assumed limited from the beginning. 

The condition (\ref{eq:eD}) for the algorithm to choose the event of type $i$ at the moment $t_s^i$ can be rewritten as
\begin{equation}
\boldsymbol{D_i}^T(\boldsymbol{x}(t_s^i) -\boldsymbol{\bar x}(t_s^i))>\omega^2
\label{eq:fc1}
\end{equation}
We now introduce a vector $\boldsymbol{F_i}$ for every type of event $i$, which is aligned with the decoding vector $D_i$, but can have an arbitrary norm $|\boldsymbol{F_i}|$: 
\begin{equation}
\boldsymbol{F_i} = \frac{|\boldsymbol{F_i}|}{\omega}\boldsymbol{D_i}
\end{equation}
In terms of this set of vectors, the condition for spike of type $i$ becomes
\begin{equation} 
\boldsymbol{F_i}^T(\boldsymbol{x}(t_s^i) -\boldsymbol{\bar x}(t_s^i))>T_i 
\label{eq:fc2}
\end{equation}
where
\begin{equation}
T_i = \omega|\boldsymbol{F_i}|
\end{equation}
We now interpret the time-dependent left-hand side of the condition (\ref{eq:fc2}) as a membrane voltage of the integrate-and-fire neuron, and $T_i$ as its firing threshold.
\begin{equation}
V_i(t) = \boldsymbol{F_i}^T(\boldsymbol{x}(t)-\boldsymbol{\bar x}(t))
\label{eq:VFe}
\end{equation}
What we called an event of type $i$ is now interpreted as a spike of neuron $i$, the decoding vectors associated to different neurons determine the matrix of synaptic connections (see below), and the vectors $\boldsymbol{F_i}$ are the feedforward vectors that describe how the $J$ components of the feedforward input are combined to obtain the input into the neuron $i$.
 
Taking the derivative of the equation (\ref{eq:VFe}) gives the time evolution of the membrane voltage:
\begin{equation}
\dot{\boldsymbol{V}}  =  \boldsymbol{F}\dot{\boldsymbol{x}} - \boldsymbol{FD}\dot{\boldsymbol r} = \boldsymbol{F}(\dot{\boldsymbol{x}} -\lambda \boldsymbol{x }+ \lambda\boldsymbol{ x}) - \boldsymbol{FD}(-\lambda \boldsymbol{r} + \boldsymbol{\rho}) = -\lambda\boldsymbol{ V} + \boldsymbol{F c}  + \boldsymbol{\Omega^\text{f}\rho}
\end{equation}
Here we have switched to matrix notation: $\boldsymbol{V}$ is an $N$-dimensional vector, $\boldsymbol{F}$ is an $N$ by $J$-dimensional matrix and $\boldsymbol{D}$ is a $J$ by $N$-dimensional matrix. We have also introduced an $N$-dimensional vector $\boldsymbol{\rho} (t)$ of delta functions at the times of the spikes of the corresponding neuron:
\begin{equation}
\rho_i(t) = \sum_{s = 1}^{N_i}\delta(t-t_s^i)
\label{eq:rho}
\end{equation}
We also introduced an $N$ by $N$ dimensional matrix of fast synaptic connections $\boldsymbol{\Omega^\text{f}}$ and a $J$-dimensional vector of feedforward inputs  $\boldsymbol{c}(t)$
\begin{align}
&\boldsymbol{\Omega^\text{f}} = -\boldsymbol{FD}\notag\\
&\boldsymbol{c}(t) = \dot{\boldsymbol{ x}}(t)+\lambda \boldsymbol{x}(t)
\end{align}

The resulting network is in direct correspondence with an infinite network with only fast connections, described in the section \ref{sec:fastOnly}. The matrix $\boldsymbol{D}$ corresponds to $\boldsymbol{D^\text{f}}$ in the section \ref{sec:fastOnly}, and the represented variable $\boldsymbol{x}(t)$ is the leaky integral of the feedforward input into the network, which we have denoted by $\boldsymbol{\hat c}(t)$ in section \ref{sec:fastOnly}.

When assuming that for any vector of norm $\omega$ in the $J$-dimensional space there is a neuron $i$ whose decoding vector $\boldsymbol{D_i}$ is arbitrarily close to it, we implicitly assume that the network has an infinite size. For a finite-size network, there is a limited set of decoding vectors, all of which still have the norm $\omega$. 
Unlike in the case of the infinite network, the norm of representation error can exceed the value of $\omega$, when there is no neuron with the decoding vector required to correct it exactly. A spike will be fired when the component of the representation error along one of the vectors $\boldsymbol{D_i}$ becomes equal to $\omega$ (equation (\ref{eq:fc1})), and only this component of the error will be removed (vector $\boldsymbol{D_i}$ will be added to the estimate $\boldsymbol{\bar x}(t)$). Choosing the best neuron in the network to fire and only when it reduces the cost function corresponds to the dynamics of vector $\boldsymbol{\hat d}$ bouncing within the firing surface as described in the section \ref{sec:fastOnly}. 

If we now assume 
\begin{equation}
\boldsymbol{F}^T\boldsymbol{F} =\gamma \boldsymbol{I}
\label{eq:Fgamma}
\end{equation}
where $\gamma$ is an arbitrary constant and $\boldsymbol{I}$ is a $J$ by $J$ identity matrix,
the cost function can be rewritten as
\begin{equation}
{\cal{L}}(t) = \frac{1}{\gamma}(\boldsymbol{x}(t)-\boldsymbol{\bar x}(t))^T\boldsymbol{F}^T\boldsymbol{F}(\boldsymbol{x}(t)-\boldsymbol{\bar x}(t)) + \omega^2n_\text{sp}(t)
\end{equation}
This form allow to express the cost function in terms of the $N$-dimensional vector of the neuronal voltages
\begin{equation}
{\cal{L}}(t) = \frac{1}{\gamma}\boldsymbol{V}^2(t) + \omega^2n_\text{sp}(t)
\end{equation} 

Note, that the assumption (\ref{eq:Fgamma}) holds for a large network with the vectors $\boldsymbol{F_i}$ drawn from an isotropic distribution.

Formulating the cost function in purely network terms demonstrates the equivalence of the notion of the feedforward-recurrent balance (see section \ref{sec:Balance}) and the efficiency of representation, unifying the encoding and the dynamical approaches.  

\section{Discussion}\label{sec:Dis}
We have shown how the structure of the feedforward input, spatial or temporal, can be incorporated into the connectivity structure of the network of integrate-and-fire neurons to allow for more efficient online representation of the input by the network's spikes. According to the main principle of the feedforward-recurrent balance, the incorporation of the input structure into the network can be interpreted as the network "having an expectation" that the future feedforward-input will obey this structure and mimicking it in the generated recurrent input. When the difference between the expected and the actual feedforward input accumulates, a spike is fired to correct for this difference and to updated the future expectation. Such a network implements the framework of predictive coding on the level of neuronal interaction. 

There are two notions of input predictability that we are concerned with. The first notion is neuron-to-neuron predictability, namely inferring the state or the input into all the neurons in the network based on the state of a particular one. The input structure that corresponds to this aspect of predictability is the low-dimensionality of the feedforward input (see (\ref{eq:Ilow})), which we often referred to as spatial structure. The network with only fast connections considered in the section \ref{sec:fastOnly} "expects" that the input changes in the low-dimensional subspace which is the image of its connectivity matrix $\boldsymbol{\Omega^\text{f}}$ and updates the states of all the neurons after each spike by the means of large short-lasting synaptic current. If the feedforward input indeed possesses this expected structure, the voltages of all the neurons are reset to zero after every population spike. The expectation about the future feedforward input in this case is that it will be zero after the spike, as is reflected in the fact that the synaptic current is short-lasting. This network, which is a slight modification of the network described in \cite{Boerlin2013}, is predictive only in the spatial sense. For the network considered in the section \ref{sec:SB}, the slowly decaying synaptic current implements the expectation that the feedforward current is smooth. To be more precise, the network expects that the feedforward current changes according to the equation $\dot{\boldsymbol{c}} = -\lambda_\text{s}\boldsymbol{c}$, where $\lambda_\text{s}$ is the slow synaptic time constant of the network. In the last presented version of the network, section \ref{sec:TS}, the network predicts that the input follows the second order differential equation $\ddot{\boldsymbol{c}}+(\lambda_1+\lambda_2)\dot{\boldsymbol c} + \lambda_1\lambda_2\boldsymbol{c} = \boldsymbol{0}$ as this is the equation obeyed by the sum of the two synaptic currents, one of which decays with the time constant $\lambda_1$ and the second with $\lambda_2$. The initial value of the input and its derivative is updated at every spike. 

In the present manuscript we discussed the optimal network that incorporates different structures present in the feedforward input. It was shown in \cite{LongSI} that in the case of the network with only fast connections, the value of the connectivity matrix that optimally incorporates the prediction of the spatial structure (low dimensionality) of the input, can be learned. Following the logic of \cite{LongSI} we propose a modified learning rule for the connectivity matrix described in the section \ref{sec:fastOnly}:
$$
d{\Omega^\text{f}_{ij}}(t) = -\varepsilon V_i\rho_j(t^-)
$$
where $V_i(t)$ is the postsynaptic voltage at the moment $t$ and $\rho_j(t^-)$ is the presynaptic neural response function defined in (\ref{eq:rho}) evaluated slightly before the moment $t$ (the synaptic weights are updated only when the presynaptic neuron spikes, the value of the postsynaptic voltage $V_i$ is evaluated after the spike), $\varepsilon$ is the learning rate. In the section \ref{sec:fastOnly} we have derived that the ratio of the firing threshold to the absolute value of the feedforward weights $\frac{T_i}{|\boldsymbol{F_i}|}$ in the optimal network is the same for all the neurons. This is achieved by equalization of the mean firing rates either with adjusting the thresholds (homeostatic threshold) or with scaling up or down all the feedforward synapses converging onto a given neuron.  
Simulations show that the proposed learning rule leads to the optimal connectivity matrix (\ref{eq:flr}) under an appropriate choice of parameters. We leave the details of the simulation beyond the scope of this paper.

We hypothesize that once the feedforward input has a corresponding temporal structure in addition to spatial and the synaptic interaction is mediated by one or more types of long-lasting currents whose effects on the postsynaptic neurons can be adjusted separately, a similar learning rule will lead to the connectivity described in sections \ref{sec:SB} - \ref{sec:DE1}. The learning rule we propose is
$$
d\Omega^{(a)}_{ij} = -\epsilon V_i(t)h^{(a)}_j(t)
$$
where $h^{(a)}_j(t)$ is the presynaptic current of type $(a)$. By different types of synaptic currents we mean different evolution of $h^{(a)}_j(t)$ after the spike. In the current work we have specified this dynamics as an exponential decay with different time constants corresponding to different types, however this was not necessary for the line of reasoning. The assumed exponential decay can be substituted with virtually any reproducible synaptic dynamics, as long as it differs sufficiently between the current types and has the range of time constants required to match the feedforward input.  

Whether this learning rule leads to the connectivity structure corresponding to expanded dimensionality of network dynamics, as described in sections \ref{sec:SB} - \ref{sec:DE1}, and how the network "chooses" the internal dimensions (the matrix $\boldsymbol{F^\text{int}}$) and the matrices $\boldsymbol{\tau}$ remains a subject of future investigation.

\begin{figure}
\includegraphics[width = 10cm]{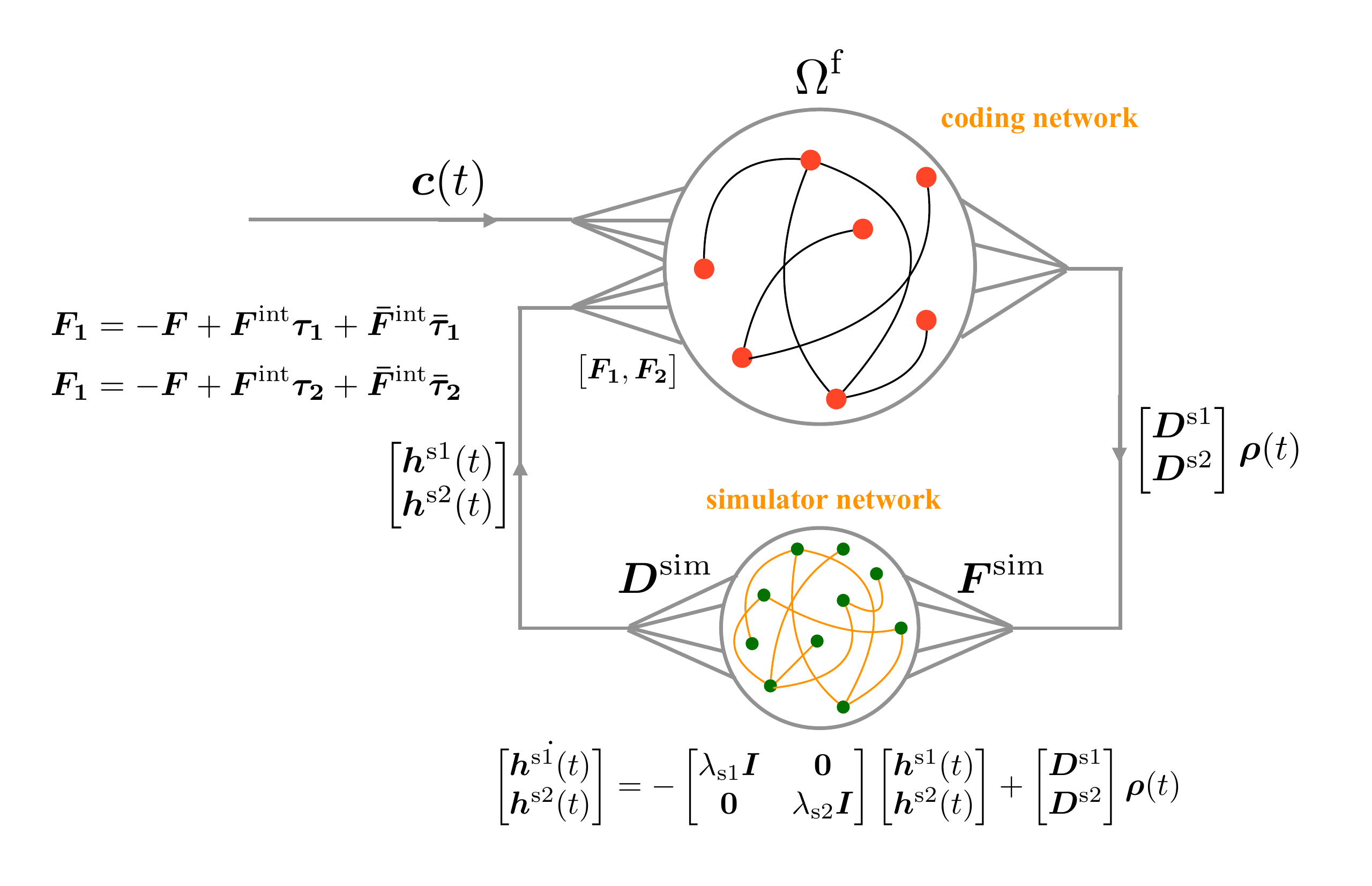}
\centering
\caption{Coding and Simulator networks}
\label{fig:CodeSim}
\end{figure}

The slow currents $\boldsymbol{h^\text{s1}}$ and $\boldsymbol{h^\text{s2}}$ of the section \ref{sec:TS}, can be generated by another network, instead of in the synapses of the network encoding the input. In figure \ref{fig:CodeSim} we show schematically the architecture of such two-subnetwork system that is equivalent to a single network described in the section \ref{sec:TS}. The coding subnetwork, which receives the feedforward input has only fast recurrent connections (see section \ref{sec:fastOnly}), while the slow synaptic input $\boldsymbol{\Omega^\text{s1}h^\text{s1}+\Omega^\text{s2}h^\text{s2}}$ is substituted by the input from the simulator network. The simulator is the network described in \cite{Boerlin2013}, whose output is the solution to a given differential equation with the source term given by the external input. This network receives a particular projection of the vector of spikes of the coding network as its input and generates in response a sustained output. In order to mimic the two synaptic currents of the section \ref{sec:TS}, the simulator network must generate a $2J$-dimensional dynamics with the first $J$ components decaying with the rate $\lambda_\text{s1}$ and the rest with the rate $\lambda_\text{s2}$ (see figure \ref{fig:CodeSim}). While mathematically the single network with slow synapses and the system of two subnetworks are equivalent, in the later the exponential decay can be substituted by any another liner or non-linear \cite{Alemi2017} dynamics. This allows to implement predictions about the feedforward input on the time scale longer than the decay of synaptic currents. When the dynamics of the simulator does not math the expected dynamics of the input, as in the example above, the dimensionality of the simulator dynamics should be higher, in order to approximate the expected evolution of the input. One could imagine, however, that when the given temporal pattern in the input persists, the recurrent connections within the simulator might change in order to reproduce the dynamics of the input \cite{LearningSlowConnections}, in which case the dimensionality of the simulator will drop to match the dimensionality of the encoded input.

\section{Acknowledgements}
LK would like to thank Christian Machens, Srdjan Ostojic, Mirjana Maras, Vincent Bouttier and Ivan Gordeli for useful discussions. 

\section{Appendix} \label{sec:App}
In this section we derive equation (\ref{eq:boundN}) the lower bound on the number of neurons $N$ in the case of expanding the dimensionality of the network dynamics in comparison with the dimensionality of the input, as described in the section \ref{sec:DE1}. In the current scenario, only a small part of the directions in the $2J-$ dimensional state space will be explored by the network. Indeed, let us consider the projection of the state vector $\boldsymbol{\hat d}(t)$ onto the input space, which is a $J-$ dimensional vector of the first $J$ components of $\boldsymbol{\hat d}(t)$, $\boldsymbol{\hat d_{1\dots J}}(t)$, and the projection of $\boldsymbol{\hat d}(t)$ onto the internal space of network dynamics, $\boldsymbol{\hat d_{J+1\dots 2J}}(t)$. The ratio of the absolute values of these projections is equal to the tangent of the angle between the state vector and the input subspace, which we call \emph{latitude} and denote by $\alpha(t)$. At the moment of a spike, the state vector $\boldsymbol{\hat d}(t)$ is equal to one of the firing vectors $\boldsymbol{\hat d ^{(i)}}$, and it follows from the firing condition (\ref{eq:ASCDE1}), that at the moment of the spike, the angle $\alpha(t^i_\text{sp})$ is determined by
\begin{equation}
\text{tan}(\alpha(t^i_\text{sp})) = \frac{|\boldsymbol{\hat d_{J+1\dots 2J}^{(i)}}|}{|\boldsymbol{\hat d_{1\dots J}^{(i)}}|} = \frac{|\boldsymbol{\tau c}(t_\text{sp}^i)|}{\lambda\omega}
\end{equation}
Assuming that the ratio $\frac{\boldsymbol{c}(t)}{\lambda}$ (which is approximately equal to $\boldsymbol{x}(t)$) is bounded by a constant of the order one, which we denote by $B$, we can write 
$$
|\text{tan}(\alpha(t^i_\text{sp}))|\leq B\frac{|\boldsymbol{\tau}|}{\omega}
$$
We assume that $\frac{|\boldsymbol{\tau}|}{\omega}$ is small, and make an approximation $\text{tan}(\alpha(t))\leq 1$.
This implies that only neurons whose firing faces are in the direction 
\begin{equation}
|\alpha| \leq B\frac{|\boldsymbol{\tau}|}{\omega}\equiv\alpha_0
\label{eq:boundAlpha}
\end{equation}
will ever fire. The reset of the space should also be covered by the firing faces of neurons to make the firing surface closed, for the case of unexpected inputs, but this coverage can have much smaller resolution. As the number of these "additional" neurons is small in comparison with the neurons whose firing faces are in the directions that satisfies (\ref{eq:boundAlpha}), we will ignore them in the following calculation. 

So, instead of covering the entire sphere of radius $\omega$ in the $2J$-dimensional space with the patches of linear size $\Delta$, we only need to cover a spherical segment determined by the constraint on the latitude (\ref{eq:boundAlpha}). We denote the area of this segment by $A_{2J,\alpha_0}$. 

$$
A_{2J,\alpha_0} = (2\alpha_0\omega)^JV_JS_J\omega^{J-1} = (2\alpha_0)^J\omega^{2J-1}V_JS_J  
$$
Here $S_J$ is the area of the unit sphere and $V_J$ is the volume of the unit ball in $J-$ dimensional space, $\alpha_0$ is defined in (\ref{eq:boundAlpha}). This area should be covered by patches of radius $\Delta$, which we approximate as $(2J-1)-$ dimensional balls. 
$$
A_{2J,\alpha_0} = N\Delta^{2J-1}V_{2J-1}
$$
Equating the two expressions for $A_{J,\alpha}$ and taking into account (\ref{eq:tauBig}) as a bound on $\Delta$, we get the bound on the number of neurons in the network
$$
N\gg \frac{(2B)^JV_JS_J}{V_{2J-1}\omega^J|\boldsymbol{\tau}|^{J-1}}\left(\frac{\lambda^\text{slow}}{\lambda}\right)^{2J-1}
$$
which is the equation (\ref{eq:boundN}). Note, that the matrix $\boldsymbol \tau$ should still be chosen to satisfy (\ref{eq:tauSmall}).

In the case of further dimensionality expansion as described in the section \ref{sec:DE2}, the above estimate changes. Let $\tau_\text{max}$ be the largest (in the absolute value) of the eigenvalues of the matrices $\boldsymbol{\tau_1}$, $\boldsymbol{\tau_2}$, $\boldsymbol{\bar\tau_1}$ and $\boldsymbol{\bar \tau_2}$. Then, the absolute value of the angle between the state vector $\boldsymbol{\hat d}(t)$ and the input space is bounded by 
$$
\alpha_{0} = \frac{\tau_\text{max}B}{\omega}
$$
The area of the corresponding spherical segment in the $3J-$dimensional space is given by
$$
A_{3J,\alpha_0} = (2\alpha_0\omega)^{2J}V_{2J}S_J\omega^{J-1} = (2\alpha_0)^{2J}\omega^{3J-1}V_{2J}S_J  
$$
the same area should be covered with $3J-1$ dimensional patches of linear size $\Delta$:
$$
A_{3J,\alpha_0} = N\Delta^{3J-1}V_{3J-1}
$$
which together with (\ref{eq:tauBig}) leads the constraint (\ref{eq:boundN3}) on the number of neurons:
 
$$
N\gg \frac{(2B)^{2J}V_{2J}S_J}{V_{3J-1}\omega^{2J}\tau_\text{max}^{J-1}}\left(\frac{\lambda^\text{slow}}{\lambda}\right)^{3J-1}
$$
\bibliographystyle{utphys}
\bibliography{ref} 
\end{document}